\documentclass[a4paper,12pt,notitlepage]{article}
\usepackage{amsfonts,amssymb,amsmath,amsthm,colonequals}
\usepackage{hyperref}
\usepackage[mathscr]{euscript}
\usepackage{bbm}
\usepackage[dvips]{graphicx}
\usepackage{makeidx}
\usepackage{url}
\usepackage{multirow}
\usepackage{ctable}
\theoremstyle{definition}
\newcommand{\beqa}{\begin{eqnarray}}
\newcommand{\eeqa}{\end{eqnarray}}
\newcommand{\beq}{\begin{equation}}
\newcommand{\eeq}{\end{equation}}

\newcommand{\tG}{\textsf{G}}
\newcommand{\tZ}{\textsf{Z}}
\newcommand{\tk}{\textsf{k}}\newcommand{\te}{\textsf{e}}\newcommand{\tf}{\textsf{f}}
\newcommand{\tP}{\textsf{P}}
\newcommand{\tH}{\textsf{H}}

\newcommand{\tth}{\textsf{h}}\newcommand{\tr}{\textsf{r}}

\newcommand{\tU}{\textsf{U}}
\newcommand{\tR}{\textsf{R}}
\newcommand{\tV}{\textsf{V}}
\newcommand{\tJ}{\textsf{J}}\newcommand{\tF}{\textsf{F}}

\newcommand{\fa}{\mathfrak{a}}\newcommand{\fb}{\mathfrak{b}}
\newcommand{\fc}{\mathfrak{c}}\newcommand{\fd}{\mathfrak{d}}

\newcommand{\fh}{\mathfrak{g}}
\newcommand{\fg}{\mathfrak{A}}

\newcommand{\cR}{\mathcal{R}}\newcommand{\cC}{\mathcal{C}}
\newcommand{\cL}{\mathcal{L}}\newcommand{\cP}{\mathcal{P}}
\newcommand{\cQ}{\mathcal{Q}}\newcommand{\cS}{\mathcal{S}}
\newcommand{\cK}{\mathcal{K}}\newcommand{\cJ}{\mathcal{J}}
\newcommand{\cB}{\mathcal{B}}
\newcommand{\bS}{\mathbb{S}}\newcommand{\bT}{\mathbb{T}}
\newcommand{\bF}{\mathbb{F}}\newcommand{\bP}{\mathbb{P}}

\def\c{\textbf{c}}\def\n{\textbf{n}}
\def\m{\textbf{m}}

\newcommand{\ket}[1]{\ensuremath{\left|\, #1\right>}}

\newcommand{\com}[2]{\ensuremath{\left[ #1\, ,\, #2\right]}}
\newcommand{\acom}[2]{\ensuremath{\left\{ #1\, ,\, #2\right\}}}

\begin{document}

\thispagestyle{empty}
\setcounter{page}{0}
\begin{flushright}\footnotesize
\texttt{HU-Mathematik-2012-10}\\
\texttt{HU-EP-12/33}\\
\texttt{AEI-2012-107}\\
\vspace{0.5cm}
\end{flushright}
\setcounter{footnote}{0}

\begin{center}
{\Large\textbf{\mathversion{bold}
The Tetrahedral Zamolodchikov Algebra and the 
$AdS_5\times S^5$ S-matrix
}\par}
\vspace{15mm}

{\sc Vladimir Mitev $^{a}$
\footnote{Present address: PRISMA Cluster of Excellence, Institut f\"ur Physik, WA THEP, Johannes Gutenberg-Universit\"at Mainz, Staudingerweg 7, 55128 Mainz, Germany}, 
Matthias Staudacher $^{a,b}$,  Zengo Tsuboi $^{a}$
\footnote{Present address: 
Laboratoire de physique th\'eorique, D\'epartement de physique 
de l'ENS, \'Ecole normale sup\'erieure, PSL Research University, 
Sorbonne Universit\'es, UPMC Univ. Paris 06, CNRS, 75005 Paris, France
} ,}\\[5mm]

{\it $^a$ Institut f\"ur Mathematik und Institut f\"ur Physik, Humboldt-Universit\"at zu Berlin\\
IRIS Haus, Zum Gro{\ss}en Windkanal 6,  12489 Berlin, Germany
}\\[5mm]

{\it $^b$ Max-Planck-Institut f\"ur Gravitationsphysik, Albert-Einstein-Institut\\
    Am M\"uhlenberg 1, 14476 Potsdam, Germany
}\\[5mm]

\texttt{vmitev$\bullet$uni-mainz.de}\\
\texttt{matthias$\bullet$aei.mpg.de}\\
\texttt{ztsuboi$\bullet$yahoo.co.jp}\\[20mm]

\textbf{Abstract}\\[2mm]
\end{center}
The S-matrix of the $AdS_5\times S^5$ string theory is a tensor product of two centrally extended su$(2|2)\ltimes \mathbb{R}^2$ S-matrices, each of which is related to the R-matrix of the Hubbard model. 
The R-matrix of the Hubbard model was first found by Shastry, who ingeniously exploited the fact that, for zero coupling, the Hubbard model can be decomposed into two XX models. In this article, we review and clarify this construction from the AdS/CFT perspective and investigate the implications this has for the $AdS_5\times S^5$ S-matrix.
\\[2mm]
{\bf Journal: Communications in Mathematical Physics (2017)}\\
(Received: 25 January 2013 / Accepted: 18 April 2017 / First Online: 23 May 2017)
\\
{\bf doi:10.1007/s00220-017-2905-y}



\tableofcontents
\addtolength{\baselineskip}{5pt}

\section{Introduction and Conclusion}

The integrability of a 1-dimensional quantum system (or a 2-dimensional classical system) is directly linked to the Yang-Baxter equation for the R-matrix of the model. Currently many of the known R-matrices are derived from symmetry arguments, using quantum affine Lie  (super)algebras. 
The most famous exception to this appears to be the R-matrix of the Hubbard model found by Shastry \cite{Shastry:1986zz} 
(cf.\ \cite{EFGKK05}). 
Interestingly, this R-matrix was derived independently in two ways. First, Shastry observed that the Hubbard model at zero coupling decomposes into two non-interacting free fermions\footnote{Also known as XX models.} and then made an Ansatz for the full R-matrix using a linear combination of (twisted) free fermion R-matrices. Secondly, in the context of the AdS/CFT correspondence, Beisert noted in \cite{Beisert:2006qh}
 that the S-matrix of the gauge theory spin chain model decomposes into a tensor product $S_1\otimes S_2$ of two matrices, each one of which is completely fixed by su$(2|2)\ltimes \mathbb{R}^2$ symmetry,
up to an overall phase which was found by other means \cite{Beisert:2006ez}.  Amazingly enough, each of the $S_i$'s coincides, see \cite{Beisert:2006qh,Martins:2007hb}, with the Hubbard R-matrix by a similarity transformation under a certain unitarity condition. For the past several years, this factor $S_i$ of the full S-matrix, which for simplicity we will call for the rest of this paper somewhat sloppily ``the AdS/CFT S-matrix'', was studied intensively in relation to the superalgebra su$(2|2)\ltimes \mathbb{R}^2$ and its Yangian (see \cite{Beisert:2005tm,hubbard-ads,Gomez:2006va} and references therein). Interestingly, prior to these discoveries, the Hubbard model had already appeared, see \cite{Rej:2005qt}, in the study of the AdS/CFT correspondence, though in a seemingly quite different context.

Originally, it was not clear that the R-matrix proposed by Shastry was a solution to the Yang-Baxter equation. This issue was clarified by Shiroishi and Wadati in \cite{Shiroishi1995a}, who not only showed that Shastry's R-matrix satisfies the Yang-Baxter equation, but later on also generalized the construction, based on a formalism due to Korepanov \cite{Korepanov:1994rc,Korepanov:1993} that he developed for Zamolodchikov's Tetrahedron algebra \cite{Zamolodchikov1980}. We find that their generalized R-matrix exactly coincides with the AdS/CFT S-matrix of \cite{Arutyunov:2006yd}, up to some similarity transformations and a condition on the parameters, without assuming the unitarity condition, as \cite{Martins:2007hb} implicitly do. Furthermore, we argue that this generalized R-matrix is for generic values of its parameters {\em not} equivalent to the AdS/CFT S-matrix. Instead, it contains it as a special case\footnote{When we compare the S-matrix in \cite{Beisert:2006qh} with the R-matrix in \cite{Shiroishi1995}, we have to correct a few minor typos in page 33 in \cite{Beisert:2006qh}: $U$ in the second and the third equations in eq.\ (4.7) has to be removed; $A_{12}$ in the last equation in eq.\ (4.10) should be $D_{12}$.}.

In the present paper, we investigate the quantum symmetry of the Hubbard model and relate it to su$(2|2)\ltimes \mathbb{R}^2$ on the AdS/CFT side. In the process, the four-layer structure of the AdS/CFT S-matrix is made apparent. While the broad lines of the relationship between the Hubbard model R-matrix and the AdS/CFT S-matrix are clear, several points either remain obscure and/or indicate possible new directions of investigation:
\begin{itemize}
\item Shastry's R-matrix has by construction an obvious two layer structure that is not at all apparent in the AdS/CFT formulation. In the present article, we strive to remedy this and to rewrite the $S_i$ matrices in a fashion to expose their two layer properties as completely as possible.
\item As mentioned, Shastry's construction was later extended by Shiroishi and Wadati based on Korepanov's formalism of the tetrahedral Zamolodchikov algebra. In the process, they obtained a generalized R-matrix, referred to here as $\tR$. We suspect that for generic values of its parameters $\tR$ is not equivalent to the AdS/CFT $S$-matrix. Rather, it seems that $\tR$ is generically less symmetric and thus strictly contains it as a special case. One of our goals is to carefully investigate the symmetries of $\tR$ and to find the precise conditions for its reduction to the $S$-matrix. 
\item Each of the two layers of the Hubbard model R-matrix enjoys  an quantum affine sl(2) symmetry. As the layers are glued together in Shastry's construction, half of this symmetry is lost, but the remaining one can then be identified with a part of the su$(2|2)\ltimes \mathbb{R}^2$ symmetry of the $S$-matrix. We carefully spell out this identification and in the process investigate and clarify the origins and importance of the R-matrix symmetries. 
\item An important, so far overlooked direction of research opened by the introduction of the tetrahedral Zamolodchikov algebra is the investigation of the three-dimensional integrable structures in the context of the AdS/CFT correspondence. Specifically, one finds a natural object in Shastry's construction, referred to as $\mathbb{S}$ in this text, which obeys the tetrahedron Zamolodchikov equation, a three-dimensional generalization of the Yang-Baxter equation. While outside the scope of the present work, we lay here the ground work for further research and hope that our results give a cue on how to reveal a hidden three-dimensional integrable structure in the context of the AdS/CFT correspondence. 
\end{itemize}
Having thus spelled out our scope and goals, we are ready to begin with the main body of our investigation.

\section{The free fermion model}

\subsection{Preliminaries}

We start our journey by writing down the free fermion model 
using oscillators
\footnote{See appendix \ref{Jordan} for formulation of the 
model without oscillators.}
 and by describing the tetrahedral Zamolodchikov algebra. For this purpose, we define the fermionic creation operators $\c_j^{\dagger}$ as well as the annihilation operators $\c_j$, where $j \in {\mathbb Z}$ labels the lattice site. They obey the canonical anti-commutation relations
\beq
\label{eq:formionicoscillators}
\acom{\c_j}{\c_k^{\dagger}}=\delta_{jk}, 
\qquad 
\acom{\c_j}{\c_k}=
\acom{\c_j^{\dagger}}{\c_k^{\dagger}}=0,
\qquad 
j,k \in {\mathbb Z}.
\eeq
It is useful to define the (bosonic) compound operators
 $\n_j\colonequals \c_j^{\dagger}\c_j$ and $\m_j\colonequals \c_j\c_j^{\dagger}$. The R-matrix of the model is a special case of the XXZ one and has the shape
\beq
\label{fermionicRmatrix}
\tR_{jk}^{\text{f}}(A)=-a\n_j\n_k-i b\n_j\m_k-ic \m_j\n_k+d\m_j\m_k+\c_j^{\dagger}\c_k+\c_k^{\dagger}\c_j,
\eeq
where the parameters satisfy the \textit{free fermion condition}, i.e. $ad-bc=1$, meaning that
\beq
\label{eq:parametrizeSL2C}
A\colonequals\left(\begin{array}{cc}a & b \\ c & d\end{array}\right) \in \text{SL}(2,\mathbb{C}).
\eeq
Thus, in a sense the R-matrix of the free fermion model has a SL$(2,\mathbb{C})$ spectral parameter, see \cite{Bazhanov:1984iw}. The Hamiltonian density of the spin chain is then obtained by choosing a curve in SL$(2,\mathbb{C})$, i.e. making $A$ depend on a parameter $u \in \mathbb{C}$ such that for $u=u_0$ the coefficients become $a=d=1$ and $b=c=0$, implying the relation $\tR^{\text{f}}_{jk}(A(u_0))=\tP_{jk}$, 
where the graded permutation is defined by 
\begin{align}
\tP_{jk}=-\n_j\n_k +\m_j\m_k+\c_j^{\dagger}\c_k+\c_k^{\dagger}\c_j. 
 \label{gr-permutation}
\end{align}
 One then constructs the transfer matrix as a supertrace of the monodromy matrix over an auxiliary space, i.e. $\tau(u)=\text{str}_{a}\left(\tR^{\text{f}}_{aN}(u)\cdots \tR^{\text{f}}_{a1}(u)\right)$ and derives the Hamiltonian  $\tH^{\text{f}}=\tau(u_0)^{-1}\frac{d}{du}\tau(u)_{|u=u_0}$. One computes that the nearest neighbor Hamiltonian is 
\beq
\label{eq:XXmodelHamiltonian}
\tH^{\text{f}}_{i,i+1}=\frac{d}{du}\check{\tR}^{\text{f}}_{i,i+1}(A(u))_{|u=u_0}.
\eeq 
Here, we introduce the notation $\check{\tR}^{\text{f}}_{jk}=\tP_{jk}\tR^{\text{f}}_{jk}$. The simplest example for such a curve is obtained by setting the parameters to $a=d=\cos u$ and $b=c=i \sin u$ leading to the purely hopping XX model Hamiltonian density $\tH^{\text{XX}}_{i,i+1}=\c^{\dagger}_i\c_{i+1}+\c^{\dagger}_{i+1}\c_{i}$. Integrability is ensured by the Yang-Baxter equation $\tR_{12}^{\text{f}}(A)\tR_{13}^{\text{f}}(B)\tR_{23}^{\text{f}}(C)=\tR_{23}^{\text{f}}(C)\tR_{13}^{\text{f}}(B)\tR_{12}^{\text{f}}(A)$ for the operator \eqref{fermionicRmatrix}, which is obeyed if the SL$(2,\mathbb{C})$ matrices fulfill the product relation $B=CA$, see \cite{Bazhanov:1984iw}. In order to have this conditions be automatic, we define a new operator $\tR^0$ as
\beq
\label{eq:defR0}
\tR^0_{jk}(A_j,A_k)\colonequals \tR^{\text{f}}_{jk}(A_kA_j^{-1}),
\eeq
so that now $\tR^0_{12}\tR^0_{13}\tR^0_{23}=\tR^0_{23}\tR^0_{13}\tR^0_{12}$ for any three SL$(2,\mathbb{C})$ matrices 
$A_{i}=\left(\begin{array}{cc}
a_{i} & b_{i} \\ c_{i} & d_{i}
\end{array}\right)$.
The above relation \eqref{eq:defR0} represents  a SL$(2,{\mathbb C})$  analogue of the difference property on the spectral parameters. Furthermore, one can easily prove the property $\tP_{23} \tR^0_{12}(A_1,A_2)\tP_{23}=\tR^0_{13}(A_1,A_2)$. Note that the right hand side is not $\tR^0_{13}(A_1,A_3)$. As a final remark, we have the inversion relation 
\beq
\label{eq:ferminveq}
\tR_{jk}^{\text{f}}(A)\tR_{kj}^{\text{f}}(A^{-1})=a d\ .
\eeq
\subsection{The quantum affine symmetry}
\label{subsec:quantumsymmetry}

In this section, we want to investigate the symmetries of the R-matrix of \eqref{fermionicRmatrix} and see the extension to which it is constrained by symmetry. A main object of concern 
is the quantum group U$_{q}(\text{sl}(2))$ at a root of unity $i$ of the deformation  parameter $q$  (see for example, chapter 7 of \cite{GRS96}).  For brevity of notation, we let $\fg$ denote the quantum group U$_q(\text{sl}(2))$, while $\hat{\fg}$ stands for its affine counterpart. We set the deformation parameter $q$ of the quantum group to $i$. The quantum group $\hat{\fg}$ is defined as generated by the operators $\tk_r$, $\te_r$ and $\tf_r$ for $r=0,1$ that for $q=i$ obey the relations
\beq
\tk_r\tk_s=\tk_s\tk_r,\quad \tk_r \te_s=-\te_s\tk_r,\quad \tk_r \tf_s=-\tf_s\tk_r,\quad \com{\te_r}{\tf_s}=\delta_{rs}\frac{\tk_r-\tk_r^{-1}}{2i},
\eeq
together with the Serre relations that we have omitted. Furthermore, we introduce the operators $\tth_r$ by the relation $\tk_r=q^{\tth_r}$, which implies
\beq
\com{\tth_r}{\te_s}=\mathbf{A}_{rs}\te_s, \quad \com{\tth_r}{\tf_s}=-\mathbf{A}_{rs}\tf_s,
\eeq
where $\mathbf{A}=(\mathbf{A}_{rs})_{0 \le r,s \le 1}$ is the Cartan matrix of affine sl(2), namely 
\beq
\mathbf{A}=\left(\begin{array}{cc}2 & -2 \\-2 & 2\end{array}\right). 
\eeq
The Hopf algebra $\hat{\fg}$ for $q=i$ has a family of two-dimensional representations (akin to nilpotent representations), denoted by $\widehat{V}_{\mu;x,y}$, which we write using the fermionic oscillators of \eqref{eq:formionicoscillators} as
\begin{align}
&\tk_0=\lambda^{-1}(\n-\m),& &\te_0=-\varphi x^{-1} \,\c^{\dagger},& &\tf_0=\varphi x\, \c,& &\tth_0=\mu-\m+\n, &\nonumber\\ &\tk_1=\lambda(\n-\m),& &\te_1=\varphi y^{-1}\,\c,&  &\tf_1=-\varphi y\,\c^{\dagger}, &  &\tth_1=-\mu+\m-\n, &
 \label{rep-fermi}
\end{align}
where $\lambda$, $\mu$, $x$ and $y$ are complex parameters, we have introduced the element $\varphi$ through the equation $\varphi^2=\frac{\lambda-\lambda^{-1}}{2i}$ and 
the lattice site index is omitted. Since $\tk_r=q^{\tth_r}$, the parameter $\lambda$ is fixed by the equation $\lambda=i^{-1-\mu}$.

At this point, a couple of remarks are in order. First, the structure of the representations only depends on the product of $x$ and $y$,  since $\widehat{V}_{\mu;x,y}$ is isomorphic to $\widehat{V}_{\mu;xy,1}$ thanks to the similarity transformation
${\mathsf J} \mapsto {\mathsf g}{\mathsf J}{\mathsf g}^{-1}$, $ {\mathsf g}=\n+y \m $, where ${\mathsf J} $ are the generators in \eqref{rep-fermi}. We choose to keep the label $y$ nonetheless, because it introduces a twisting by an inner automorphism that will be useful later on. Note further that the representation $\widehat{V}_{\mu;xy,1}$ is based on a homogeneous gradation of $\hat{\mathfrak g}$ and that the squares of some generators are  central elements, specifically we have non-trivial central elements ${\mathsf k}_{0}^2=\lambda^{-2}, {\mathsf k}_{1}^2=\lambda^{2}$ and trivial ones ${\mathsf e}_{r}^2={\mathsf f}_{r}^2=0$.

Second, the $\hat{\fg}$ modules $\widehat{V}_{\mu;x,y}$  are also representations of the Hopf subalgebra $\fg\subset \hat{\fg}$ that is generated by the operators $\tk_0$, $\te_0$, $\tf_0$ and $\tth_0$. We  shall refer to them as $V_{\mu}$ since they do not depend on the parameters $x$ or $y$. Lastly, there is an ambiguity in the sign of $\varphi$ due to the square root in its definition. This does not lead to different representations however, because one can change the sign of $\varphi$ via the inner automorphism $\tJ\mapsto \tk_0\tJ\tk_0^{-1}$. 

In defining the coproduct, we introduce two additional operators, the first being the grading operator $\tF\colonequals (-1)^{\n}=\m-\n$, while the second one is an additional central element $\tZ$.  At first we define the coproduct for the diagonal elements
\beq
\label{eq:Delta1}
\Delta(\tk_r)=\tk_r\otimes\tk_r, \quad \Delta(\tZ)=\tZ\otimes \tZ, \quad \Delta(\tF)=\tF\otimes\tF,\quad \Delta(\tth_r)=\tth_r\otimes \mathbbm{1}+\mathbbm{1}\otimes \tth_r,
\eeq
while for the non-diagonal one we set
\begin{align}
\label{eq:Delta2}
&\Delta(\te_0)=\te_0\otimes \tZ+\tk_0 \tF \otimes \te_0,& &\Delta(\tf_0)=\tf_0\otimes \tk_0^{-1}\tZ^{-1}+
 \tF \otimes \tf_0,& \nonumber\\
&\Delta(\te_1)=\te_1\otimes \mathbbm{1} +\tZ\tk_1 \tF \otimes\te_1,& &\Delta(\tf_1)=\tf_1\otimes \tk_1^{-1}+\tZ^{-1} \tF \otimes\tf_1.& 
\end{align}
The presence of the grading operator $\tF$ can be explained by the fermionic nature of the operators\footnote{It can also be explained by first defining the coproduct in the usual way and then performing a Jordan-Wigner transformation.} $\te_r$ and $\tf_r$. Were we to represent the $\hat{\fg}$ generators with matrices instead of oscillators, we would not need $\tF$. Note that the tensor product is graded, so that for instance $(\mathbbm{1}\otimes \tf_0)(\te_0\otimes \mathbbm{1})=-\te_0\otimes \tf_0$. The central element $\tZ$ is used to twist the coproduct\footnote{Let us consider an automorphism of $\hat{\fg}$ defined 
by $[{\mathsf B}_{i},{\mathsf e}_{j}]=\delta_{ij} {\mathsf e}_{j}$, 
 $[{\mathsf B}_{i},{\mathsf f}_{j}]=-\delta_{ij} {\mathsf f}_{j}$, 
 $[{\mathsf B}_{i},{\mathsf h}_{j}]=0$, $i,j=0,1$. 
Then we can remove the central element  ${\mathsf Z}$ from the 
co-product \eqref{eq:Delta2} by the following map
\begin{align}
\Delta({\mathsf X}) \mapsto 
(1 \otimes {\mathsf Z})^{-{\mathsf B}_{0} \otimes 1} 
({\mathsf Z} \otimes 1)^{-1 \otimes {\mathsf B}_{1}} 
\Delta({\mathsf X})
({\mathsf Z} \otimes 1)^{1 \otimes {\mathsf B}_{1}} 
(1 \otimes {\mathsf Z})^{{\mathsf B}_{0} \otimes 1} 
\quad 
\text{for} 
\quad 
{\mathsf X} \in \hat{\mathfrak{g}}. \nonumber
\end{align}
Moreover this map can be realized as a composition of a change of 
the basis and a Reshetikhin-twist 
\cite{Reshetikhin:1990ep} for the R-matrix (see the discussion in section 6 of \cite{Beisert:2011wq}). 
This type of map may be useful to connect the AdS/CFT S-matrix with 
Shiroishi-Wadati's generalized Hubbard R-matrix. 
In fact, as we will see later, the 
central element is not free but effectively $\pm i$ for the AdS/CFT S-matrix. 
}
 and we impose on it the requirement that its eigenvalue $z$ be the same in every space. This twisting will become necessary in section \ref{sec:symmetries} in order to connect the quantum symmetry to the S-matrix of  \eqref{fermionicRmatrix}. 

We now look for an intertwiner $\tr^0_{12}$ acting on the space $\widehat{V}_{\mu_1;x_1,y_1}\otimes \widehat{V}_{\mu_2;x_2,y_2}$ subject to the condition 
\beq
\label{eq:invquantumaffine}
\Delta^{\prime}(\textsf{J})\tr^0_{12}=\tr^0_{12}\Delta(\textsf{J})
  \qquad \text{for} \quad \forall \textsf{J}\in \hat{\fg}\ ,
\eeq
where\footnote{Here $\sigma (X \otimes Y):=(-1)^{p(X)p(Y)}Y \otimes X $ for any elements $X,Y $of the algebra ($p(X)$ is the grading of $X$, 
$p(\c)=p(\c^{\dagger})=1, p(\n)=0$).} $\Delta^{\prime}:=\sigma \circ \Delta $  is the opposite coproduct to the one of \eqref{eq:Delta1} and  \eqref{eq:Delta2}. For example, one has $\Delta^{\prime}(\te_0)=\tZ \otimes\te_0 +\te_0\otimes \tk_0 \tF$. The solution to the symmetry constraints \eqref{eq:invquantumaffine} is unique up to normalization and can be written explicitly as 
\begin{align}
\label{eq:rquantumsym}
&\tr^0_{12}=(x_1 y_1\lambda_1\lambda_2-x_2y_2)\n_1\n_2+z^{-1}(x_2 y_2\lambda_1-x_1y_1\lambda_2)\n_1\m_2+z(x_2 y_2\lambda_2-x_1y_1\lambda_1)\m_1\n_2\nonumber\\&+(x_1 y_1-x_2y_2\lambda_1\lambda_2)\m_1\m_2-\sqrt{(\lambda_1-\lambda_1^{-1})(\lambda_2-\lambda_2^{-1})}\left(x_1y_2\lambda_2\c_2^{\dagger}\c_1+x_2y_1\lambda_1\c_1^{\dagger}\c_2\right).
\end{align}
The operator $\tr_{12}^{0}$ obeys the Yang-Baxter equation automatically since the tensor product of three $\hat{\fg}$ representations $\widehat{V}_{\mu_r;x_r,y_r}$ is generically irreducible. This means that there can be only one  $\hat{\fg}$-invariant intertwiner and since both $\tr^0_{12}\tr^0_{13}\tr^0_{23}$ and $\tr^0_{23}\tr^0_{13}\tr^0_{12}$ are by construction invariant intertwiners, they must be equal up to a multiplicative constant that is easily seen to be one. One remarks that the representation labels $\mu_r$ only enter \eqref{eq:rquantumsym} through the $\lambda_r$, but they will play a more explicit role later on. Furthermore, the $y_{i}$ dependence of \eqref{eq:rquantumsym} can be removed by a similarity transformation
\begin{align}
 \label{eq:map-gauge}
\tr^0_{12} \mapsto 
{\mathsf g}_{1}{\mathsf g}_{2}\tr^0_{12}
 {\mathsf g}_{2}^{-1}{\mathsf g}_{1}^{-1}, 
\qquad {\mathsf g}_{i}=\n_{i}+y_{i} \m_{i} , 
\quad i=1,2. 
\end{align}
 and the rescaling of the spectral parameter $x_{i} \mapsto  x_{i}y_{i}^{-1}$.

If we now define the operators $\tG_s\colonequals\m_s+\sqrt{\frac{y_s}{x_s}\lambda_s}\n_s$, then it turns out that $\tr^0_{12}$ can be identified with the free fermion operator $\tR^0$ of \eqref{eq:defR0} in the following way
\beq
\label{eq:identr0}
\tR_{12}^0=-\frac{1}{\sqrt{(\lambda_1-\lambda_1^{-1})(\lambda_2-\lambda_2^{-1})x_1 y_1 x_2 y_2 \lambda_1\lambda_2}}\tG_1^{-1}\tG_2^{-1} \tr^0_{12}\tG_1\tG_2\ ,
\eeq
provided that we relate the parameters as
\beq
\label{eq:changevarquant}
a_r=\sqrt{\frac{(x_ry_r)^{-1}\lambda_r}{\lambda_r-\lambda_r^{-1}}},\quad b_r=\frac{1}{iz}\sqrt{\frac{x_ry_r\lambda_r^{-1}}{\lambda_r-\lambda_r^{-1}}}, \quad c_r=i z\sqrt{\frac{(x_ry_r)^{-1}\lambda_r^{-1}}{\lambda_r-\lambda_r^{-1}}}, \quad d_r=\sqrt{\frac{x_ry_r\lambda_r}{\lambda_r-\lambda_r^{-1}}}.
\eeq
It is here that the need for the introduction of the central operator $\tZ$ becomes apparent. We remind that $z$ is the eigenvalue of the $\tZ$ and that it is a global number, i.e. it does not depend on the lattice index. We observe that with the identification \eqref{eq:changevarquant} we get the relation
\begin{align}
a_r b_r=-z^{-2} c_rd_r\quad 
\text{for all} \quad r. 
 \label{symcon}
\end{align}
 This means that $z$ is necessary in order to cover the full space of parameters SL$(2,\mathbb{C})$ of the operator $\tR_{12}^0$. Lastly, as we have noted before, the role of the labels $y_r$ is to twist the $\textsf{J}_1$ generators relative to the $\textsf{J}_0$ ones. When such a twist is not necessary, we will simply set $y_r=x_r$, which corresponds to the principal gradation of $\hat{\mathfrak g}$. For convenience, in the next section we summarize all the relations between the free fermion variables and the quantum group variables in table \ref{table1}.

\subsection{The tetrahedral Zamolodchikov algebra}
In the previous sections, we described how the XX R-matrix is uniquely fixed by the requirement \eqref{eq:invquantumaffine} of invariance under $\hat{\fg}$. It will however turn out to be useful to relax that condition. Let us consider the quantum group $\fg \subset \hat{\fg}$ that is generated by the elements  $\tk_0$, $\te_0$, $\tf_0$ and $\tth$. We denote a representation given by restricting the generators in \eqref{rep-fermi} to the ones for  $\fg$ as\footnote{This is a two-dimensional irreducible $\fg $ module generated by the highest weight vector $v$ defined by $ \te_{0}v=0$, $\tth v=(\mu +1)v$, $\tk_{0} v=\lambda^{-1} v $.} $V_{\mu}$. Although \eqref{rep-fermi} depends on the spectral parameter $x$, this representation is essentially independent of it since it can be removed by a similarity transformation. 

If we now consider \eqref{eq:invquantumaffine} 
 only for the quantum group $\fg\subset \hat{\fg}$, 
 then the space of solutions to the equation 
\beq
\Delta^{\prime}(\tJ)\tr_{12}=\tr_{12}\Delta(\tJ)\quad \forall \tJ\in \fg\eeq is two-dimensional, since the tensor product of two $\fg$ modules $V_{\mu_r}$ decomposes into two irreducible pieces\footnote{Unless $\mu_1+\mu_2 =0$ mod 2. } as 
\beq
V_{\mu_1}\otimes V_{\mu_2}\cong V_{\mu_1+\mu_2+1}
\oplus V_{\mu_1+\mu_2-1}.
\eeq
A basis for this space is given by the set $\{\tr_{12}^0,\tr_{12}^1\}$. The first of these operators is the operator of the previous section $\tr_{12}^0$, which is the solution \eqref{eq:rquantumsym} to the affine invariance equations \eqref{eq:invquantumaffine} for the representations $\widehat{V}_{\mu_1;x_1,x_1}\otimes \widehat{V}_{\mu_2;x_2,x_2}$. The second one, we call  $\tr_{12}^1$ and it is also a solution of \eqref{eq:invquantumaffine}, but for the representations $\widehat{V}_{\mu_1;x_1,x_1}\otimes \widehat{V}_{\mu_2;x_2,-x_2}$, i.e. the action on the second factor has been twisted by a minus sign\footnote{Let us consider an automorphism ${\mathsf X} \mapsto e^{i \pi {\mathsf B}_{1}}{\mathsf X} e^{-i \pi {\mathsf B}_{1}}$ for any ${\mathsf X} \in \fg $. This induces a map $\widehat{V}_{\mu;x,y} \mapsto \widehat{V}_{\mu;x,-y}$ 
on the representation. Thus the above twisting can be interpreted as a consequence of this map.}. 

A rescaling of both of these operators by the same factor in \eqref{eq:identr0}, followed by a similarity transformation as in \eqref{eq:identr0} with $\tG_r=\m_r+\sqrt{\lambda_r}\n_r$ and a change of variables as in \eqref{eq:changevarquant} but with $y_r=x_r$ leads to the basis of operators $\{\tR^0, \tR^1\}$  with
\beq
\tR^0_{12}(A_1,A_2):=\tR^{\text{f}}_{12}(A_2A_1^{-1}), \quad \tR^1_{12}(A_1,A_2):=\tR^{\text{f}}_{12}(A_2\sigma^{z}A_1^{-1}\sigma^{z})(\n_2-\m_2),  
 \label{r0r1}
\eeq
where $\sigma^{z}=\text{diag}(1,-1)$ and the $A_i$ are elements of SL$(2,\mathbb{C})$. We want to use these two operators to describe the space of $\fg$-invariant intertwiners on the tensor product of three modules $V_{\mu_r}$, which decomposes generically as
\beq
\label{eq:decomp3}
V_{\mu_1}\otimes V_{\mu_2}\otimes V_{\mu_3}\cong 
V_{\mu_1+\mu_2+\mu_3 +2}\oplus 2 V_{\mu_1+\mu_2+\mu_3}\oplus V_{\mu_1+\mu_2+\mu_3-2}.
\eeq
Thanks to the invariance properties of the $\tR^0$ and $\tR^1$ it is clear that the $16$ operators $\tR_{12}^{\alpha}\tR_{13}^{\beta}\tR_{23}^{\gamma}$ and $\tR_{23}^{\alpha}\tR_{13}^{\beta}\tR_{12}^{\gamma}$ for $\alpha,\beta,\gamma \in \{0,1\}$ 
are all $\fg$-invariant. However,  \eqref{eq:decomp3} tells us that there at most six of them  can be linearly independent, since  the dimension of the space of such invariant intertwiners is by Schur's lemma equal to $1^2+2^2+1^2=6$. The relationships between the various intertwiners is described by two equations: 
the defining relations of the tetrahedral Zamolodchikov algebra 
and the linear dependence equations. In order to write them down, it turns out to be useful to perform a change of basis to  ``light-cone'' operators $\tR^{\pm}\colonequals \frac{1}{2}(\tR^0\pm \tR^1)$, explicitly written using oscillators as
\beqa
\label{eq:defRpm}
\tR^{+}_{jk}(A_j,A_k)&=&(a_k\n_j+ic_k\m_j)(-d_j\n_k+ib_j\m_k)+\c_j^{\dagger}\c_k\nonumber\\
\tR^{-}_{jk}(A_j,A_k)&=&(b_k\n_j+id_k\m_j)(c_j\n_k-ia_j\m_k)+\c_k^{\dagger}\c_j\ .
\eeqa
These operators satisfy the following set of relations 
\beq
\label{eq:tetrahedronZamalgebra}
\tR^{\alpha}_{23}\tR^{\beta}_{13}\tR^{\gamma}_{12}=\sum_{\bar{\alpha},\bar{\beta},\bar{\gamma}=\pm} \bS^{\alpha\beta\gamma}_{\bar{\alpha}\bar{\beta}\bar{\gamma}}\, \tR^{\bar{\gamma}}_{12}\tR^{\bar{\beta}}_{13}\tR^{\bar{\alpha}}_{23}\ ,
\eeq
where the coefficients $\bS^{\alpha\beta\gamma}_{\bar{\alpha}\bar{\beta}\bar{\gamma}}$ are given by 
(cf.\ \cite{Shiroishi1995})
\begin{align}
\label{eq:defcoeffS}
\bS^{\alpha\,\beta\,\gamma}_{\bar{\alpha}\,\bar{\beta}\,\bar{\gamma}}&=\delta^{\alpha\,\beta\,\gamma}_{\bar{\alpha}\,\bar{\beta}\,\bar{\gamma}}\qquad  \text{ for } (\alpha,\beta,\gamma)\notin\left\{(+,-,+),(-,+,-)\right\}\nonumber\\
\bS^{+-+}_{\bar{\alpha}\,\bar{\beta}\,\bar{\gamma}}&= \delta^{-+-}_{\bar{\alpha}\,\bar{\beta}\,\bar{\gamma}}+\bF^{23}_1\left(\frac{b_3d_3}{b_2d_2}\delta^{++-}_{\bar{\alpha}\,\bar{\beta}\,\bar{\gamma}}-\frac{a_3c_3}{a_2c_2}\delta^{--+}_{\bar{\alpha}\,\bar{\beta}\,\bar{\gamma}}\right)-\bF^{21}_3\left(\frac{b_1d_1}{b_2d_2}\delta^{+--}_{\bar{\alpha}\,\bar{\beta}\,\bar{\gamma}}-\frac{a_1c_1}{a_2c_2}\delta^{-++}_{\bar{\alpha}\,\bar{\beta}\,\bar{\gamma}}\right),\nonumber\\
\bS^{-+-}_{\bar{\alpha}\,\bar{\beta}\,\bar{\gamma}}&= \delta^{+-+}_{\bar{\alpha}\,\bar{\beta}\,\bar{\gamma}}-\bF^{23}_1\left(\frac{b_3d_3}{b_2d_2}\delta^{++-}_{\bar{\alpha}\,\bar{\beta}\,\bar{\gamma}}-\frac{a_3c_3}{a_2c_2}\delta^{--+}_{\bar{\alpha}\,\bar{\beta}\,\bar{\gamma}}\right)+\bF^{21}_3\left(\frac{b_1d_1}{b_2d_2}\delta^{+--}_{\bar{\alpha}\,\bar{\beta}\,\bar{\gamma}}-\frac{a_1c_1}{a_2c_2}\delta^{-++}_{\bar{\alpha}\,\bar{\beta}\,\bar{\gamma}}\right).
\end{align}
Here, we use the abbreviations $\delta^{\alpha\,\beta\,\gamma}_{\bar{\alpha}\,\bar{\beta}\,\bar{\gamma}}\colonequals \delta^{\alpha}_{\bar{\alpha}}\delta^{\beta}_{\bar{\beta}}\delta^{\gamma}_{\bar{\gamma}}$ and $\bF^{jk}_i\colonequals \frac{a_i d_i-a_jd_j}{a_i d_i-a_kd_k}$.
An algebra generated by six operators 
$\{\tR^{\alpha}_{ij}\}$ satisfying the 
relations of the form 
\eqref{eq:tetrahedronZamalgebra} 
supplemented by the linear dependence relations and the consistency condition 
(which we mention later in \eqref{eq:lindependence} and \eqref{eq:almostTZE}) is called 
the {\em tetrahedral Zamolodchikov algebra}
\footnote{The equation 
\eqref{eq:tetrahedronZamalgebra} itself is also called the tetrahedral Zamolodchikov algebra 
in some references in the literature. 
Korepanov defined this for the 
asymmetric free-fermion 8 vertex model. In this case, \eqref{eq:lindependence} becomes void since the corresponding 
eight operators $\{ \tR^{\alpha}_{12}\tR^{\beta}_{13}\tR^{\gamma}_{23}\}$ 
are independent. He also found an explicit solution of the equation \eqref{eq:tetrahedronZamalgebra} 
for the symmetric free-fermion 8 vertex model. However, his solution does not coincide 
with \eqref{eq:defcoeffS} even in the limit to the symmetric free-fermion 6 vertex model.}  
\cite{Korepanov:1993}. 
 Formally, the matrix elements $\bS^{\alpha\,\beta\,\gamma}_{\bar{\alpha}\,\bar{\beta}\,\bar{\gamma}}$ define an endomorphism $\bS$ of the space of $\fg$-invariant operators. As follows from \eqref{eq:decomp3}, the eight operators $\tR^{\alpha}_{12}\tR^{\beta}_{13}\tR^{\gamma}_{23}$ are not linearly independent, but span a six-dimensional space. One finds that they are subject to the following two linear dependence relations
\beq
\label{eq:lindependence}
\sum_{\alpha,\beta,\gamma=\pm}\bT^{(i)}_{\alpha \beta \gamma}\tR^{\alpha}_{12}\tR^{\beta}_{13}\tR^{\gamma}_{23}=0\quad \text{ for } i=1,2.
\eeq
Here, the coefficients are explicitly
\beq
\bT^{(1)}_{+++}=\frac{a_1b_3}{b_1a_3}, \quad \bT^{(1)}_{--+}=\frac{b_1c_2}{a_1d_2}, \quad \bT^{(1)}_{+--}=\frac{d_2a_3}{c_2 b_3}, \quad \bT^{(1)}_{+-+}=1.
\eeq
The $\bT^{(2)}$ coefficients are obtained by making the formal parameter exchanges $a_r\leftrightarrow c_r$ and $b_r\leftrightarrow d_r$, while the remaining ones follow from the relation $\bT^{(i)}_{-\alpha,-\beta,-\gamma}=(\bT^{(i)}_{\alpha \beta\gamma})^{-1}$. Using both equations \eqref{eq:lindependence}, we can write the eight vectors $\tR^{\alpha}_{12}\tR^{\beta}_{13}\tR^{\gamma}_{23}$ in terms of six linear independent vectors, which we do not specify. We denote the $6 \times 8$ matrix for this change of basis as $\bP$. Note that, due to \eqref{eq:lindependence}, the coefficients appearing in \eqref{eq:defcoeffS} are not unique. In fact, one has a 16 parameter freedom since any $\bS^{\prime}$ with
\beq
\label{eq:transformedStensor}
\big(\bS^{\prime}\big)^{\alpha\,\beta\,\gamma}_{\bar{\alpha}\,\bar{\beta}\,\bar{\gamma}}=\bS^{\alpha\,\beta\,\gamma}_{\bar{\alpha}\,\bar{\beta}\,\bar{\gamma}}+
\sum_{i=1}^{2}
c^{\alpha\,\beta\,\gamma}_{(i)}\bT^{(i)}_{\bar{\gamma}\,\bar{\beta}\,\bar{\alpha}}
\eeq
will obey 
the defining relations of the tetrahedral Zamolodchikov algebra 
\eqref{eq:tetrahedronZamalgebra} for any value of the complex parameters $c^{\alpha\,\beta\,\gamma}_{(i)}$.

We end this section by considering the application of the  
the relations \eqref{eq:tetrahedronZamalgebra} to the product of six operators $\tR^{\pm}$, namely 
the consistency condition of the tetrahedral Zamolodchikov algebra. 
Specifically, one finds that there are two a priori inequivalent ways of transforming the product $\tR^a_{34}\tR^b_{24}\tR^c_{14}\tR^d_{23}\tR^e_{13}\tR^f_{12}$ into a linear combination of products of six $\tR^{\pm}_{ij}$ with the reverse lattice order. We make apparent the fact that the coefficients in \eqref{eq:defcoeffS} depend explicitly on the free fermion parameters $\{a_{k},b_{k},c_{k},d_{k} \}_{k=1}^{3}$ by writing them as  
$(\bS_{123})^{\alpha\beta\gamma}_{\bar{\alpha}\bar{\beta}\bar{\gamma}}$. We define analogously the coefficients of
$\bS_{124}$, $\bS_{134}$ and $\bS_{234}$, 
and find:
\begin{multline}
\label{eq:almostTZE}
\left(\big(\mathbb{S}_{123}\big)^{def}_{d^{\prime}e^{\prime}f^{\prime}}\big(\mathbb{S}_{124}\big)^{bcf^{\prime}}_{b^{\prime}c^{\prime}f^{\prime\prime}}
\big(\mathbb{S}_{134}\big)^{ac^{\prime}e^{\prime}}_{a^{\prime}c^{\prime\prime}e^{\prime\prime}}
\big(\mathbb{S}_{234}\big)^{a^{\prime}b^{\prime}d^{\prime}}_{a^{\prime\prime}b^{\prime\prime}d^{\prime\prime}}\right.\\\left.-\big(\mathbb{S}_{234}\big)^{abd}_{a^{\prime}b^{\prime}d^{\prime}}\big(\mathbb{S}_{134}\big)^{a^{\prime}ce}_{a^{\prime\prime}c^{\prime}e^{\prime}}
\big(\mathbb{S}_{124}\big)^{b^{\prime}c^{\prime}f}_{b^{\prime\prime}c^{\prime\prime}f^{\prime}}
\big(\mathbb{S}_{123}\big)^{d^{\prime}e^{\prime}f^{\prime}}_{d^{\prime\prime}e^{\prime\prime}f^{\prime\prime}}\right)\tR^{f^{\prime \prime}}_{12}\tR^{e^{\prime \prime}}_{13}\tR^{d^{\prime \prime}}_{23}\tR^{c^{\prime \prime}}_{14}\tR^{b^{\prime \prime}}_{24}\tR^{a^{\prime \prime}}_{34}=0
\end{multline}
where the Einstein summation convention applies. However, because of the linear dependence equations of \eqref{eq:lindependence}, we cannot simply set the coefficients in the sum of \eqref{eq:almostTZE} to zero. One needs to use a transformation of the kind \eqref{eq:transformedStensor} in order to obtain a tensor $\mathbb{S}^{\prime}_{ijk}$ that obeys the Zamolodchikov's 
\textit{tetrahedron equations}\footnote{This equation should be interpreted as an equation in
 $\mathrm{End}(({\mathbb C}^{2})^{\otimes 6})$.
Let us introduce $2\times 2$ matrix units $e_{ab}$ 
whose $(i,j)$ elements 
are given by  
$(e_{ab})_{ij}=\delta_{ai}\delta_{bj}$, and define 
$e_{ab}^{(12)}=e_{ab}\otimes {\mathbbm 1}^{\otimes 5}$, 
$e_{ab}^{(13)}= {\mathbbm 1} \otimes e_{ab}\otimes {\mathbbm 1}^{\otimes 4}$, 
$e_{ab}^{(23)}={ \mathbbm 1}^{\otimes 2} \otimes e_{ab}\otimes {\mathbbm 1}^{\otimes 3}$, 
$e_{ab}^{(14)}={ \mathbbm 1}^{\otimes 3} \otimes e_{ab}\otimes {\mathbbm 1}^{\otimes 2}$, 
$e_{ab}^{(24)}={\mathbbm 1}^{\otimes 4} \otimes e_{ab}\otimes {\mathbbm 1}$, 
$e_{ab}^{(34)}={\mathbbm 1}^{\otimes 5} \otimes e_{ab}$, where 
${\mathbbm 1}=e_{11}+e_{22}$. 
Then the tensors in \eqref{eq:tetrahedoron} 
are defined by 
$\bS_{ijk}^{\prime}=\sum_{a,b,c,d,e,f=\pm} 
(\bS_{ijk}^{\prime})^{abc}_{def}
e_{cf}^{(ij)} e_{be}^{(ik)} e_{ad}^{(jk)}$. 
}
:
\beq
\mathbb{S}_{123}^{\prime}\mathbb{S}_{124}^{\prime}
\mathbb{S}_{134}^{\prime}\mathbb{S}_{234}^{\prime}
=\mathbb{S}_{234}^{\prime}\mathbb{S}_{134}^{\prime}
\mathbb{S}_{124}^{\prime}\mathbb{S}_{123}^{\prime}.
\label{eq:tetrahedoron}
\eeq
Unfortunately, such a solution seems to be known only 
in the symmetric case for which $a_k=d_k=\cos u_k$ and   $b_k=c_k=-i \sin u_k$, 
see \cite{Korepanov:1993,Horibe:1994yx}. Whether a solution exists for the general case seems to be an open problem. 

\section{The Shastry-Shiroishi-Wadati R-matrix}
\label{sec:ShastryShiroishyWadatiRmatrix}

Here, we wish to review the construction of the R-matrix of the 
one-dimensional Hubbard model and generalizations thereof. We start first by taking a look at the Hamilton operator for the one-dimensional Hubbard model, which reads
\beq
\tH^{\text{Hub}}=
-\sum_{j=1}^N\sum_{\sigma=\uparrow , \downarrow}\left(
\c^{\dagger}_{j+1,\sigma}\c_{j,\sigma}+\c_{j,\sigma}^{\dagger}\c_{j+1,\sigma}\right)+\frac{U}{4}\sum_{j=1}^N\left(\m_{j,\uparrow}-\n_{j,\uparrow}\right)\left(\m_{j,\downarrow}-\n_{j,\downarrow}\right), 
\eeq
where we introduced two copies of the fermionic oscillators 
 $\c_{j,\sigma}$ and $\c_{j,\sigma}^{\dagger}$  for $\sigma=\uparrow,\downarrow$ that satisfy 
\begin{align}
\label{eq:formionicoscillators2}
\acom{\c_{j,\sigma}}{\c_{k,\tau}^{\dagger}}=\delta_{jk}\delta_{\sigma \tau}, 
\quad 
\acom{\c_{j,\sigma}}{\c_{k,\tau}}=
\acom{\c_{j,\sigma}^{\dagger}}{\c_{k,\tau}^{\dagger}}=0, 
\quad \sigma, \tau =  \uparrow \text{or} \downarrow ,
\end{align}
and defined the bosonic compound operators 
 $\n_{j,\sigma}\colonequals \c_{j,\sigma}^{\dagger}\c_{j,\sigma}$ and  $\m_{j,\sigma} \colonequals \c_{j,\sigma} \c_{j,\sigma}^{\dagger}$. 
The non-negative integer $N$ is the number of 
one-dimensional lattice sites. We impose the periodic boundary conditions $\c_{N+1,\sigma}^{\dagger}=\c_{1,\sigma}^{\dagger}$, 
$\c_{N+1,\sigma}=\c_{1,\sigma}$, 
$\sigma =  \uparrow , \downarrow $. 
Here, the number $U$ is the coupling constant. 
One sees that at $U=0$, the spin chain decomposes into two non-interacting XX models, described by the oscillators $\c_{j,\uparrow}$ and $\c_{j,\downarrow}$ respectively. The R-matrix of this product model is of course simply the product $\tR^{0}_{jk,\uparrow} \tR^{0}_{jk,\downarrow}$ of the R matrices of the individual models\footnote{We denote the operators ${\mathcal O}$ obtained by replacing the fermionic oscillators  $\c_{j},\c_{j}^{\dagger}$ therein with  $\c_{j,\alpha},\c_{j,\alpha}^{\dagger}$ for $\alpha=\uparrow$ or $ \downarrow$ by ${\mathcal O}_{,\alpha}$.}. This observation led Shastry in \cite{Shastry:1986zz} into making an Ansatz for all $U$, later generalized by Shiroishi and Wadati \cite{Shiroishi1995}, of the following form
\beq
\label{eq:SSWRmatrix}
\tR_{jk}\colonequals\tR^0_{jk,\uparrow} \tR^0_{jk,\downarrow}+\alpha_{jk} \left(\tR^0_{jk,\uparrow} \tR^1_{jk,\downarrow}+\tR^1_{jk,\uparrow} \tR^0_{jk,\downarrow}\right)+\beta_{jk}\tR^1_{jk,\uparrow} \tR^1_{jk,\downarrow},
\eeq
where $j,k \in \{1,2,\dots, N\}$, and the unknown coefficients $\alpha_{jk},\beta_{jk} \in \mathbb{C}$ are to be determined by the requirements that \eqref{eq:SSWRmatrix} obeys the Yang-Baxter equation:
\begin{align}
\tR_{12}\tR_{13}\tR_{23}=\tR_{23}\tR_{13}\tR_{12}
\end{align}
and that they both vanish in the free fermion limit $U\rightarrow 0$. One can formulate these constraints better by rewriting the operator $\tR_{jk}$ in the basis $\tR^{\pm}$ of \eqref{eq:defRpm} as  $\tR_{jk}= \sum_{\alpha,\beta=\pm} \gamma_{jk;\alpha\beta}\tR^{\alpha}_{jk,\uparrow} \tR^{\beta}_{jk,\downarrow}$. Then, making use of the defining relations of 
the tetrahedral Zamolodchikov algebra  \eqref{eq:tetrahedronZamalgebra} and of the linear dependence \eqref{eq:lindependence} leads to the set of equations\footnote{Let us introduce unit row vectors $v_{a}$ which satisfy $v_{c} e_{ab}=\delta_{ca}v_{b}$. Then the tensors and vectors in \eqref{eq:YBETetrahedron} are defined by $\bS=\sum_{a,b,c,d,e,f=\pm} 
\bS^{abc}_{def}e_{cf}\otimes e_{be}\otimes e_{ad}$, $\gamma_{12}=\sum_{a,b=\pm}\gamma_{12;ab}v_{a} \otimes {\mathbbm 1} \otimes {\mathbbm 1} \otimes v_{b} \otimes {\mathbbm 1}
 \otimes {\mathbbm 1}$, $\gamma_{13}=\sum_{a,b=\pm}\gamma_{13;ab} {\mathbbm 1}
  \otimes v_{a} \otimes {\mathbbm 1} \otimes {\mathbbm 1} 
   \otimes v_{b} \otimes {\mathbbm 1}$, 
  $\gamma_{23}=\sum_{a,b=\pm}\gamma_{23;ab}{\mathbbm 1} \otimes  {\mathbbm 1}
   \otimes v_{a} \otimes {\mathbbm 1} \otimes {\mathbbm 1}
    \otimes v_{b} $. Then \eqref{eq:YBETetrahedron} 
    correspond to the equations (4.3)-(4.7) in \cite{Shiroishi1995}.
}
\beq
\label{eq:YBETetrahedron}
\gamma_{12}\gamma_{13}\gamma_{23}
\left(\mathbbm{1}\otimes \mathbbm{1}-\bS\otimes \bS\right)
\bP\otimes \bP=0\quad \text{ and } \quad \gamma_{jk;\alpha\beta}(U=0)= 1
\quad \text{for} \quad \forall j,k,\alpha,\beta.
\eeq
These equations can be simplified if one assumes certain symmetries of the coefficients, for instance $\alpha_{jk}=0$ in the original application to the Hubbard model. The most general known solution to \eqref{eq:YBETetrahedron} has been found\footnote{To be precise, the solution in \cite{Shiroishi1995} is written in terms of some matrices rather than fermionic oscillators. 
See \eqref{eq:generalRmatrixSpin} in appendix \ref{connection-SW} for the 
 exact relation between \eqref{eq:generalRmatrixfermionic} and 
the original R-matrix by Shiroishi and Wadati \cite{Shiroishi1995}. 
In appendix \ref{connection-Bei}, 
we also establish the exact connection to the original 
work by Beisert \cite{Beisert:2006qh}.} 
in \cite{Shiroishi1995} and leads to an $\tR_{jk}$ matrix that depends, not on just one constant $U$, but rather on two complex parameters that we name $\Theta$ and $\Xi$. Explicitly, the most general non-trivial solution has the form
\beq
\label{eq:generalRmatrixfermionic}
\tR_{jk}=\frac{v_k+\frac{b_j c_j}{a_j d_j} v_j}{\frac{b_k c_k}{a_k d_k} v_k+v_j}\left(\frac{a_j b_k}{b_j a_k}\tR^+_{jk,\uparrow} \tR^+_{jk,\downarrow}+\frac{d_j c_k}{c_j d_k}\tR^-_{jk,\uparrow} \tR^-_{jk,\downarrow}\right)+\tR^+_{jk,\uparrow} \tR^-_{jk,\downarrow}+\tR^-_{jk,\uparrow} \tR^+_{jk,\downarrow}\ ,
\eeq
where we have suppressed the dependence of the operators\footnote{From now on, we often use a shorthand notation on indices for  any operators ${\mathcal O}$: ${\mathcal O}_{\uparrow\downarrow}$ denotes ${\mathcal O}_{\uparrow}$ or ${\mathcal O}_{\downarrow}$. When we need to write the matrix dependence of $\tR_{jk}$, we use the notation $\tR_{jk}(A_{j},A_{k})$.} $\tR^{\pm}_{jk,\uparrow\downarrow}$ on the matrices $A_j$ and $A_k$. We note though that the $\uparrow$ and $\downarrow$ layers both use the same matrices. The new variables $v_j$, referred to as \textit{gluing parameters}, are not free but depend on the constants $\Theta$ and $\Xi$ through the \textit{gluing conditions}:
\beq
\label{eq:gluingcond}
i\frac{\Theta^2}{a_k d_k} v_k-i\frac{\Xi^2}{b_kc_k}v_k^{-1}=1
\quad \text{ for } \quad k=1,2.
\eeq
The obvious two-layer structure of the operator $\tR$ leads to a number of interesting relations, which we elaborate on in the appendix \ref{double-free}.
The R-matrix of the Hubbard model
\footnote{See appendix \ref{connection-Sha} for the exact connection to the original work 
by Shastry.}
 is a special case of \eqref{eq:generalRmatrixfermionic} for which the global parameters take the values  
\begin{align}
\Theta^2=-\Xi^2=-i U^{-1} ,
 \label{Hu-limit1}
\end{align}
the SL$(2,\mathbb{C})$ parameters are specialized to 
\begin{align}
a_j=d_j=\cos u_j, \qquad b_j=c_j=-i\sin u_j 
  \label{Hu-limit2}
\end{align} and we replace the gluing parameters with $h_j$ via 
\begin{align}
v_j=e^{2h_j}\cot u_j. 
  \label{Hu-limit3}
\end{align}
This then reduces the gluing equation of \eqref{eq:gluingcond} to the well-known formula 
\begin{align}
\sinh{2 h_j}=\frac{U}{4}\sin{2u_j}, 
  \label{Hu-limit4}
\end{align}
which relates the extra parameters appearing in the R-matrix to the coupling constant $U$ and the spectral parameter. One should note that in general, the solutions to \eqref{eq:gluingcond} involve elliptic functions (see \cite{Janik:2006dc} for more details on the use of elliptic functions in the context of the AdS/CFT S-matrix\footnote{We thank V.Kazakov and especially A.Zabrodin for interesting discussions on the elliptic parametrization of the  the AdS/CFT S-matrix.}). This does not however mean that the model is elliptic in the usual sense, since \eqref{eq:generalRmatrixfermionic} lacks a difference property.

We can make several observations regarding the symmetries of the R-matrix \eqref{eq:generalRmatrixfermionic}. First, one notes that the R-matrix is invariant under the exchange of the two layers  $\uparrow$ and $\downarrow$. Second, one sees that part of the quantum symmetry of the free fermion building blocks survives.  Specifically, we still have a $\fg\oplus \fg$ quantum group symmetry\footnote{This  corresponds to \eqref{eq:invquantumaffine} at each layer.} 
generated by the elements $\tk_{0,\uparrow \downarrow}, \te_{0,\uparrow \downarrow},\tf_{0,\uparrow \downarrow}$ and $\tth_{0,\uparrow \downarrow}$, namely
\beq
\label{eq:remainingQuantumSymmetryR}
\left(\tG^{-1}\Delta^{\prime}(\tJ)\tG\right)\tR_{12}=\tR_{12}\left(\tG^{-1}\Delta(\tJ)\tG\right)\text{ with } \tJ\in\{\te_{0,\uparrow\downarrow},\tf_{0,\uparrow\downarrow},\tk_{0,\uparrow\downarrow},\tth_{0,\uparrow\downarrow}\}.
\eeq
Here we have defined the operator $\tG:=\tG_{1,\uparrow}\tG_{1,\downarrow}\tG_{2,\uparrow}\tG_{2,\downarrow}$, see \eqref{eq:identr0}.  Thus, the two-layer form of the R-matrix makes apparent its invariance under the four fermionic generators $\te_{0,\uparrow \downarrow},\tf_{0,\uparrow \downarrow}$. In establishing the connection \eqref{eq:remainingQuantumSymmetryR}, it is useful to keep in mind to relationships between the free fermion variables and the quantum group variables, summarized in table \ref{table1}. In the next section, we shall see how the quantum symmetry is related to the larger algebra  su$(2|2)\ltimes \mathbb{R}^2$, which contains eight fermionic generators.

\begin{table}
\begin{center}
\begin{tabular}[h]{lr}
\toprule[0.15 em] \textbf{Quantum group variables}&\textbf{Free fermion variables}\\
\toprule[0.08 em] $x_r$, $\lambda_r$, $z$, $\varphi_r$, $\mu_r$ &  $a_r$, $b_r$, $c_r$, $d_r$\\
$\lambda_r=i^{-1-\mu_r}$ &$a_rd_r-b_rc_r=1$\\
 $\varphi_r^2=\frac{\lambda_r-\lambda_r^{-1}}{2i}$ & $a_r=x_r^{-1}\sqrt{\frac{\lambda_r}{\lambda_r-\lambda_r^{-1}}}$\\
$\lambda_r^2=\frac{a_rd_r}{b_rc_r}$&$b_r=\frac{x_r}{iz} \sqrt{\frac{\lambda_r^{-1}}{\lambda_r-\lambda_r^{-1}}}$\\
$x_r^2=\frac{d_r}{a_r}$ &  $c_r=iz x_r^{-1}\sqrt{\frac{\lambda_r^{-1}}{\lambda_r-\lambda_r^{-1}}}$\\
$z^2=-\frac{c_rd_r}{a_rb_r}$ &$d_r=x_r \sqrt{\frac{\lambda_r}{\lambda_r-\lambda_r^{-1}}}$ \\
\toprule[0.08 em] \multicolumn{2}{c}{\textbf{Gluing parameters and variables: $\Theta$, $\Xi$, $v_r$}}\\\toprule[0.08 em]
$\Theta^2\lambda_r^{-1}v_r-\Xi^2\lambda_rv_r^{-1}=
-\frac{1}{2}\varphi_r^{-2}$ &  
$i\frac{\Theta^2}{a_r d_r} v_r-i\frac{\Xi^2}{b_rc_r}v_r^{-1}=1$
\end{tabular}
\end{center}
\caption{The connections between the free fermion variables and the quantum group variables as well as the gluing conditions in both languages.}
\label{table1}
\end{table}

\section{Symmetries}
\label{sec:symmetries}

In this section, we want to investigate the symmetry properties of the operator \eqref{eq:generalRmatrixfermionic}. First, we shall connect the results of \cite{Shiroishi1995a} to those of \cite{Beisert:2006qh,Beisert:2005tm, Arutyunov:2006yd} by defining the superalgebra su$(2|2)\ltimes \mathbb{R}^2$ and showing that it is sufficient to determine the R-matrix in a certain limit. In a further part, we connect the quantum symmetry of the XX model to the superalgebra. 

\subsection{Realization of the centrally extended superalgebra}

As written for example in \cite{Beisert:2006qh}, the superalgebra $\fh_0\colonequals \text{psu}(2|2)$ is spanned by the six\footnote{We have $\cL^{1}_{\phantom{1}1}=-\cL^{2}_{\phantom{2}2}$ as well as  $\cR^{1}_{\phantom{1}1}=-\cR^{2}_{\phantom{2}2}$.} even generators $\cL^{\alpha}_{\phantom{\alpha}\beta}$ and $\cR^a_{\phantom{a}b}$ and the eight odd generators $\cQ^{\alpha}_{\phantom{\alpha}a}$ and $\cS^{a}_{\phantom{a}\alpha}$, 
where the indices $\alpha,\beta, a,b$ run over $\{1,2\} $. 
One can extend this superalgebra by adding three even central charges $\cC$, $\cP$ and $\cK$ to get $\fh\colonequals\text{su}(2|2)\ltimes \mathbb{R}^2$. The commutation relations for $\fh$ can be summarized as follows:
\begin{align}
&\com{\cL^{\alpha}_{\phantom{\alpha}\beta}}{\cL^{\gamma}_{\phantom{\gamma}\xi  }}=
\delta^{\gamma}_{\beta}\cL^{\alpha}_{\phantom{\alpha} \xi  }
-\delta^{\alpha}_{\xi}\cL^{\gamma}_{\phantom{\gamma}\beta  },
&&
\com{\cR^{a}_{\phantom{a}b}}{\cR^{c}_{\phantom{c}d}}
=\delta^{c}_{b}\cR^{a}_{\phantom{a}d}-
\delta^{a}_d\cR^{c}_{\phantom{c}b},
&\nonumber \\
&\com{\cL^{\alpha}_{\phantom{\alpha}\beta}}{\cQ^{\gamma}_{\phantom{\gamma}b  }}=
\delta^{\gamma}_{\beta}\cQ^{\alpha}_{\phantom{\alpha}b  }
-\frac{1}{2}\delta^{\alpha}_{\beta}\cQ^{\gamma}_{\phantom{\gamma}b  },
&&
\com{\cL^{\alpha}_{\phantom{\alpha}\beta}}{\cS^{a}_{\phantom{a}\gamma}} = -\delta_{\gamma}^{\alpha}\cS^{a}_{\phantom{a}\beta}
+\frac{1}{2}\delta^{\alpha}_{\beta}\cS^{a}_{\phantom{a}\gamma},
&\nonumber \\
&\com{\cR^{a}_{\phantom{a}b}}{\cS^{c}_{\phantom{c}\beta}}
=\delta^{c}_{b}\cS^{a}_{\phantom{a}\beta}-
\frac{1}{2}\delta^{a}_b\cS^{c}_{\phantom{c}\beta},
&&\com{\cR^{a}_{\phantom{a}b}}{\cQ^{\alpha}_{\phantom{\alpha}c  }}
=-\delta_{c}^{a}\cQ^{\alpha}_{\phantom{\alpha}b  }
+\frac{1}{2}\delta^{a}_b
\cQ^{\alpha}_{\phantom{\alpha}c  },
&\nonumber\\
&\big\{\cQ^{\alpha}_{\phantom{\alpha}a}\,,\,\cQ^{\beta}_{\phantom{\beta}b}\big\}
=\epsilon^{\alpha\beta}\epsilon_{ab}\cP,
&& \acom{\cS^{a}_{\phantom{a}\alpha}}{\cS^{b}_{\phantom{b}\beta}}=
\epsilon^{ab}\epsilon_{\alpha\beta}\cK,
&\nonumber\\
&\acom{\cQ^{\alpha}_{\phantom{\alpha}a}}{\cS^{b}_{\phantom{b}\beta}}=
\delta^{b}_{a}\cL^{\alpha}_{\phantom{\alpha}\beta}+
\delta^{\alpha}_{\beta}\cR^b_{\phantom{b}a}+
\delta^b_a\delta^{\alpha}_{\beta}\cC.&
\end{align}
In the above relations, $\epsilon$ is the standard antisymmetric tensor with $\epsilon^{12}=\epsilon_{12}=1$. Using two fermionic oscillators $\c_{\uparrow}$ and $\c_{\downarrow}$ we can easily obtain an important class of four-dimensional representations of $\fh$. The even part of the algebra is represented as
\begin{align}
&\cR^{1}_{\phantom{1}1}=-\cR^{2}_{\phantom{2}2}=\frac{1}{2}\left(1-\n_{\uparrow}-\n_{\downarrow}\right),&&\cR^{1}_{\phantom{1}2}= \c_{\downarrow}\c_{\uparrow},&  &\cR^{2}_{\phantom{2}1}=\c_{\uparrow}^{\dagger}\c_{\downarrow}^{\dagger},&\nonumber\\
&\cL^{1}_{\phantom{1}1}=-\cL^{2}_{\phantom{2}2}=\frac{1}{2}\left(\n_{\uparrow}-\n_{\downarrow}\right),& &\cL^{1}_{\phantom{1}2}=\c_{\uparrow}^{\dagger}\c_{\downarrow},& &\cL^{2}_{\phantom{2}1}=\c_{\downarrow}^{\dagger}\c_{\uparrow},&
\label{evengen}
\end{align}
while the odd one takes the form
\begin{align}
\label{eq:fermionicgenerators}
&\cQ^{1}_{\phantom{1}1}=(\fa\m_{\downarrow}+\fb\n_{\downarrow})\c_{\uparrow}^{\dagger},&  \qquad &\cQ^{2}_{\phantom{2}1}=(\fa\m_{\uparrow}+\fb\n_{\uparrow})\c_{\downarrow}^{\dagger},&\nonumber\\
&\cQ^{1}_{\phantom{1}2}=-(\fb\m_{\uparrow}+\fa\n_{\uparrow})\c_{\downarrow},& \qquad &\cQ^{2}_{\phantom{2}2}=(\fb\m_{\downarrow}+\fa\n_{\downarrow})\c_{\uparrow},&\nonumber\\
&\cS^{1}_{\phantom{1}1}=(\fd\m_{\downarrow}+\fc\n_{\downarrow})\c_{\uparrow},&  \qquad &\cS^{2}_{\phantom{2}1}=-(\fc\m_{\uparrow}+\fd\n_{\uparrow})\c_{\downarrow}^{\dagger},&\nonumber\\
&\cS^{1}_{\phantom{1}2}=(\fd\m_{\uparrow}+\fc\n_{\uparrow})\c_{\downarrow},& \qquad &\cS^{2}_{\phantom{2}2}=(\fc\m_{\downarrow}+\fd\n_{\downarrow})\c_{\uparrow}^{\dagger},&
\end{align}
where $\fa$, $\fb$, $\fc$ and $\fd$ are a priori free complex parameters. Closure of the algebra requires that 
$\fa \fd-\fb \fc=1$ and  the central charges take the values
\beq
\label{eq:centralchargesparameters}
\cC=\frac{\fa \fd+\fb \fc}{2}, \quad \cP=\fa\fb, \quad \cK=\fc \fd.
\eeq
We group the central charges in the vector $\vec{\cC}=(\cC,\cP,\cK)$ and denote by $V(\vec{\cC})$ the four-dimensional representation of $\fh$ generated by the operators \eqref{eq:fermionicgenerators} on a vacuum $\ket{0}$ annihilated by $\c_{\uparrow\downarrow}$. The condition $\fa \fd-\fb \fc=1$ translates to $\cC^2-\cP\cK=\frac{1}{4}$ for the central charges. The outer automorphism group of the centrally extended superalgebra $\fh$ is isomorphic to SL$(2,\mathbb{C})$ and it acts on the representations $V(\vec{\cC})$ by sending
\beq
\label{eq:transfD}
 \mathfrak{D}\mapsto  \mathfrak{D}\mathfrak{E}, \text{ for } \mathfrak{D}\colonequals \left(\begin{array}{cc}\fa& \fb\\\fc& \fd\end{array}\right)  \text{ and }   \mathfrak{E}\in\text{SL}(2,\mathbb{C}). 
\eeq
It turns out that we can represent one of the generators\footnote{This is the one extra generator that turns sl$(2|2)$ into gl$(2|2)$.} of this outer automorphism group, $\cB$, using the fermionic oscillators as follows
\beq
\cB=\n_{\uparrow}\n_{\downarrow}+\m_{\uparrow}\m_{\downarrow}.
\eeq
The operator $\cB$ commutes with the even generators and generates the transformation \eqref{eq:transfD} with $\mathfrak{E}=\text{diag}(e^{-i\phi},e^{i\phi})$,  
where $\phi \in \mathbb{C}$, namely 
$e^{i\phi \cB} {\mathcal J} (\mathfrak{D}) e^{-i\phi \cB}=
 {\mathcal J} (\mathfrak{D}\mathfrak{E})$, 
where $ {\mathcal J} (\mathfrak{D}) $ is the generator in 
 \eqref{evengen}-\eqref{eq:fermionicgenerators} as 
a function of the matrix $\mathfrak{D}$ in \eqref{eq:transfD}. 

One of the original motivations for the investigations of the relationships between the AdS/CFT S-matrix and the Hubbard model R-matrix was to determine in what way the $\text{SL}(2,\mathbb{C})$ group appearing here as the outer automorphism group of the algebra $\fh$ is related to the $\text{SL}(2,\mathbb{C})$ group that plays an important role for the free fermion R-matrix, see \eqref{eq:parametrizeSL2C}. This will be the subject of the subsections \ref{sec:invfertr} and \ref{sec:connquantumsym}.

\subsection{Invariance of the \texorpdfstring{$\tR$}{\textbf{R}} matrix under bosonic transformations}
\label{sec:invboscond}

One can check easily that the operator \eqref{eq:generalRmatrixfermionic} obeys the following invariance equations\footnote{We denote the operators ${\mathcal O}$ acting on the lattice site $j$, namely ${\mathcal O}$ obtained by replacing the fermionic oscillators $\c_{\alpha},\c_{\alpha}^{\dagger}$ therein with  $\c_{j,\alpha},\c_{j,\alpha}^{\dagger}$ for $\alpha=\uparrow$ or $ \downarrow$ by $({\mathcal O})_{j}$.}. 
\beq
\label{eq:Rinvariancecom}
\com{\tR_{jk}}{\big(\cL^{\alpha}_{\phantom{\alpha}\beta}\big)_j+\big(\cL^{\alpha}_{\phantom{\alpha}\beta}\big)_k}=0, \quad \com{\tR_{jk}}{\big(\cR^1_{\phantom{1}1}\big)_j+\big(\cR^1_{\phantom{1}1}\big)_k}=0\ ,
\eeq
independently of the gluing conditions \eqref{eq:gluingcond}. The invariance under the raising or lowering generators of the other sl$(2)$ is tricky and leads to anticommutation relations. One finds that
\beq
\label{eq:Rinvarianceanticom}
\acom{\tR_{jk}}{\frac{b_j}{c_j}\big(\cR^1_{\phantom{1}2}\big)_j-\frac{b_k}{c_k}\big(\cR^1_{\phantom{1}2}\big)_k}=0, \quad \acom{\tR_{jk}}{\frac{c_j}{b_j}\big(\cR^2_{\phantom{2}1}\big)_j-\frac{c_k}{b_k}\big(\cR^2_{\phantom{2}1}\big)_k}=0,
\eeq
if and only if the following equation that we refer to as \textit{symmetry conditions} is fulfilled:
\beq
\label{eq:symmetryconditions}
a_kb_k=c_k d_k \qquad \forall k.
\eeq 
We remark that this corresponds to the condition that 
the eigenvalue $z$ of the central element ${\mathsf Z}$ 
coincides with $i$ or $-i$ (see \eqref{symcon}). 
Note that  the Hubbard model satisfies this condition \eqref{eq:symmetryconditions}, since it obeys the even stronger condition $a_r=d_r$ and $b_r=c_r$.
We wish to rewrite the invariance conditions in a way that would make them more apparent and that would make the symmetry of the Hamilton operator transparent while strengthening the connection to the works \cite{Beisert:2006qh, Arutyunov:2006yd}. We start by defining the gauge transformation matrices $\tU_r$ and $\tV_r$
\begin{align}
\label{eq:simtranfUV}
&\tU_r\colonequals \m_{r,\uparrow}\m_{r,\downarrow}+t_r \m_{r,\uparrow}\n_{r,\downarrow}+t_r \n_{r,\uparrow}\m_{r,\downarrow}+\frac{c_r}{b_r}\n_{r,\uparrow}\n_{r,\downarrow},\nonumber\\
&\tV_r\colonequals( \m_{r,\uparrow}-i  \n_{r,\uparrow})( \m_{r,\downarrow}-i  \n_{r,\downarrow})\ ,
\end{align}
where we have introduces a new parameter $t_r \in {\mathbb C}$, 
that will remain unconstrained for now. We use this operators to perform a similarity transformation in order to get the new operator $\check{\tR}^{\prime}_{jk}\colonequals\tP_{jk} \left(\tU_j^{-1}\tU_k^{-1} \tR_{jk}\tU_j \tU_k\right)$ where the two-layer permutation operator is defined as $\tP_{12}\colonequals \tP_{12,\uparrow} \tP_{12,\downarrow}$. This new operator obeys the same invariance conditions \eqref{eq:Rinvariancecom} with $\cL^{\alpha}_{\phantom{\alpha}\beta}$ and $\cR^1_{\phantom{1}1}$ as $\tR$ but the remaining ones,  specifically \eqref{eq:Rinvarianceanticom}, turn to
\beq
\label{eq:Rcheckinvariance1}
\com{\check{\tR}^{\prime}_{jk}}{\big(\cR^1_{\phantom{1}2}\big)_j-\big(\cR^1_{\phantom{1}2}\big)_k}=0, \quad \com{\check{\tR}^{\prime}_{jk}}{\big(\cR^2_{\phantom{2}1}\big)_j-\big(\cR^2_{\phantom{2}1}\big)_k}=0.
\eeq
We emphasize again that the above equations are true only if the symmetry conditions \eqref{eq:symmetryconditions} are obeyed. Furthermore, the operator $\check{\tR}^{\prime}$ inherits from the $\tR$ the property
\beq
\check{\tR}^{\prime}_{12}(A_{2},A_{3})
\check{\tR}^{\prime}_{23}(A_{1},A_{3})
\check{\tR}^{\prime}_{12}(A_{1},A_{2})
=
\check{\tR}^{\prime}_{23}(A_{1},A_{2})
\check{\tR}^{\prime}_{12}(A_{1},A_{3})
\check{\tR}^{\prime}_{23}(A_{2},A_{3}).
\eeq
We want to use this residual symmetry in order to get rid of the minus signs in \eqref{eq:Rcheckinvariance1}. For this purpose, we perform a similarity transformation of the $\check{\tR}^{\prime}$, leading to the definition of the operator $\check{\tR}$ as
\beq
\label{eq:Rcheckfermionic}
\check{\tR}_{j,j+1}=\left\{\begin{array}{ll}\tV_j^{-1}\check{\tR}_{j,j+1}^{\prime} \tV_j &\text{ if } j\ \text{ is even}\\\tV_{j+1}^{-1}\check{\tR}_{j,j+1}^{\prime} \tV_{j+1} &\text{ if } j\ \text{ is odd}\end{array}\right.
\eeq
Finally, we get the invariance equations that we want, namely
\beq
\label{eq:Rcheckinvariance2}
\com{\check{\tR}_{j,j+1}}{\big(\cL^{\alpha}_{\phantom{\alpha}\beta}\big)_j+\big(\cL^{\alpha}_{\phantom{\alpha}\beta}\big)_{j+1}}=0, \quad \com{\check{\tR}_{j,j+1}}{\big(\cR^a_{\phantom{a}b}\big)_j+
\big(\cR^a_{\phantom{a}b}\big)_{j+1}}=0 ,
\eeq
for all $\alpha,\beta,a,b \in \{1,2\}$ and for all $j \in {\mathbb Z}$. Thus, we get the proper invariance under the bosonic symmetry provided that we look at the operator $\check{\tR}$ between nearest neighbor sites. Since the nearest neighbor Hamiltonian of the integrable system associated to $\tR$ is simply obtained, following the algebraic Bethe Ansatz, as in \eqref{eq:XXmodelHamiltonian} for the XX model by taking the logarithmic derivative of the operator $\check{\tR}$, the $\text{sl}(2)\oplus \text{sl}(2)$ symmetry is guaranteed. The only possible problem comes from boundary conditions. If we put the model on a periodic chain with a odd number of lattice sites, then \eqref{eq:Rcheckinvariance2} will not be true for the last site and the global $\cR$ symmetry will be broken. In the Hubbard model language, the $\cL$ symmetry corresponds to the spin SU$(2)$ symmetry, while the $\cR$ one is the $\eta$-pairing SU$(2)$ symmetry. It was noted in \cite{Umeno1998} that the R-matrix of the Hubbard model in the fermionic formulation anticommutes with the $\eta$-pairing ladder generators, which reflected the well established fact that this second SU$(2)$ symmetry was only present for chains of even length. 

\subsection{Invariance under fermionic transformations}
\label{sec:invfertr}

The invariance of the operator $\check{\tR}_{12}$ under fermionic transformations introduced in \eqref{eq:fermionicgenerators} is quite subtle. The action of $\check{\tR}$ transforms the central charges in a non-linear way and one is furthermore faced with the challenge of finding the relations between the $\fh$ labels $\fa$, $\fb$, $\fc$ and $\fd$ on the one hand and the free fermion variables $a$, $b$, $c$ and $d$ that enter formula \eqref{eq:Rcheckfermionic} on the other. The picture is further muddied by the presence of the gluing parameters $v_i$ and of the extra coefficients $t_i$ appearing in the similarity transformations. 

The identifications are done as follows. Given two SL$(2,\mathbb{C})$ matrices $A_j$ written as in \eqref{eq:parametrizeSL2C} with parameters that obey the symmetry condition \eqref{eq:symmetryconditions}, we define the following sets of matrices
\beq
\label{eq:defofthematricesBC}
\mathfrak{B}_j\colonequals\sqrt{\Theta \Xi}e^{-i\frac{\pi}{4}}\left(\begin{array}{cc} \frac{\Theta}{\Xi} \frac{c_j }{a_j d_j}\frac{v_j}{t_j} &\frac{\Xi}{\Theta} \frac{t_j}{c_j v_j}\\ -\frac{1}{b_j t_j}& -\frac{t_j}{c_j} \end{array}\right),\quad \mathfrak{C}_j\colonequals\left(\begin{array}{cc}\frac{c_j}{a_j}&0\\0&\frac{a_j}{c_j}\end{array}\right).
\eeq
The determinants of these matrices is one because of the gluing conditions \eqref{eq:gluingcond}. 
Thus these are also elements of SL$(2,\mathbb{C})$. 
Setting then $\mathfrak{D}_1=\mathfrak{C}_2\mathfrak{B}_1$,  $\mathfrak{D}_2=\mathfrak{B}_2$, $\mathfrak{D}_2^{\prime}=\mathfrak{C}_1\mathfrak{B}_2$ and $\mathfrak{D}_1^{\prime}=\mathfrak{B}_1$, we find the following invariance condition, valid for all non-zero values of the parameters $t_r$:
\beq
\label{eq:ferminvariance}
\check{\tR}_{12}(A_1,A_2)\left[\cJ_1(\mathfrak{D}_1)+\cJ_2(\mathfrak{D}_2)\right]
=\left[\cJ_1(\mathfrak{D}_2^{\prime})+\cJ_2(\mathfrak{D}_1^{\prime})\right]\check{\tR}_{12}(A_1,A_2)\ ,
\eeq
where $\cJ_{i}({\mathfrak D})$ is any of the fermionic operator in \eqref{eq:fermionicgenerators} 
acting on the lattice site $i \in \{1,2\}$. 
Here we regard this as a function of the matrix ${\mathfrak D}$ in  \eqref{eq:transfD}. 
 The action of the fermionic generators can be understood in the following way. On the left hand side of \eqref{eq:ferminvariance}, the $\cJ_1+\cJ_2$ act on the tensor product of two $\fh$ representations $V(\vec{\cC_1})\otimes V(\vec{\cC_2})$. In our notation, it is understood that $\cJ_i$ always acts on the $i^{\text{th}}$ factor in a tensor product. However, on the right hand side of \eqref{eq:ferminvariance}, the $\cJ_1+\cJ_2$ act on the tensor product $V(\vec{\cC_2^{\prime}})\otimes V(\vec{\cC_1^{\prime}})$ because the operator $\check{\tR}_{12}$ has exchanged the labels of the central charges, meaning that  $\check{\tR}_{12}$ is to be seen as an operator
\beq
\check{\tR}_{12}:V(\vec{\cC_1})\otimes V(\vec{\cC_2})\rightarrow V(\vec{\cC_2^{\prime}})\otimes V(\vec{\cC_1^{\prime}}). 
\eeq
We remind that the central charges $\vec{\cC_r}$, respectively $\vec{\cC_r}^{\prime}$ are related to the matrices $\mathfrak{D}_r$, respectively $\mathfrak{D}_r^{\prime}$ via \eqref{eq:centralchargesparameters}.
From the explicit expression of \eqref{eq:defofthematricesBC}, we find the relations:
\begin{align}
&\cC_1=i\frac{\Theta^2 v_1}{a_1d_1}-\frac{1}{2},& &\cP_1=\frac{ \Theta\Xi}{i a_1d_1}\left(\frac{c_2}{a_2}\right)^2,& &\cK_1=\frac{ \Theta\Xi}{ib_1c_1}\left(\frac{a_2}{c_2}\right)^2,&\nonumber\\
&\cC_2=i\frac{\Theta^2 v_2}{a_2d_2}-\frac{1}{2},& &\cP_2=\frac{ \Theta\Xi}{i a_2d_2},& &\cK_2=\frac{ \Theta\Xi}{i b_2c_2}.&
\end{align}
From the above, we find the transformation law relating the central charges of the representations before and after the action of $\check{\tR}$:
\beq
\label{eq:centralchargestransf}
\cC_i^{\prime}=\cC_i, \quad \cP_i^{\prime}=\cK_i\frac{\cP_1+\cP_2}{\cK_1+\cK_2}, \quad \cK_i^{\prime}=\cP_i\frac{\cK_1+\cK_2}{\cP_1+\cP_2}, 
\quad i=1,2,
\eeq
in complete agreement with (3.6) of \cite{Beisert:2006qh}. 

At this point, we are able to relate the variables of the free fermion formulation of the $\check{\tR}_{12}$ operator with the variables commonly found in the AdS/CFT literature.  One can explicitly check that the operator $\tP_{12}\check{\tR}_{12}$ is to be 
identified
\footnote{See appendix \ref{connection-AFZ} for the exact connection to the original work 
by Arutyunov, Frolov and Zamaklar.}
 with the AdS/CFT infinite volume S-matrix $S_{12}(p_1,p_2)$ in the string basis  provided in \cite{Arutyunov:2006yd}. Comparisons of \eqref{eq:ferminvariance} with the fermionic invariance equations  (4.11) of the same article leads to the following expressions for the matrices $\mathfrak{D}_j$ :
\beq
\label{eq:exprDAdSCFTvariables}
\mathfrak{D}_1=
\sqrt{g}\left(\begin{array}{cc}
e^{i\frac{p_2}{2}}\eta_1 & e^{i\frac{p_2}{2}} \frac{i}{\eta_1}\left(\frac{x_1^+}{x_1^-}-1\right)\\
-e^{-i\frac{p_2}{2}}\frac{\eta_1}{x_1^+} & e^{-i\frac{p_2}{2}}\frac{x_1^+}{i\eta_1}\left(1-\frac{x_1^-}{x_1^+}\right)
\end{array}\right), 
\quad 
\mathfrak{D}_2=
\sqrt{g}\left(\begin{array}{cc}
\eta_2 & \frac{i}{\eta_2}\left(\frac{x_2^+}{x_2^-}-1\right)\\
-\frac{\eta_2}{x_2^+ }& \frac{x_2^+}{i\eta_2}\left(1-\frac{x_2^-}{x_2^+}\right)
\end{array}\right).
\eeq
The matrices $\mathfrak{D}_j^{\prime}$ are then obtained by performing the formal exchange of indices $1\leftrightarrow 2$ on the right hand sides of the above equations. Hence, comparing \eqref{eq:exprDAdSCFTvariables} with \eqref{eq:defofthematricesBC} allows us to find the relations between the free fermion/gluing variables entering $\check{\tR}$ and the AdS/CFT variables. Specifically, we get
\beq
\label{eq:variablesidentifications}
x_k^+=\frac{\Theta}{\Xi}\frac{b_kc_k}{a_k d_k}v_k, \quad x_k^-=\frac{\Theta}{\Xi}v_k,\quad e^{i\frac{p_k}{2}}=\frac{c_k}{a_k}, \quad  \eta_k=e^{-i\frac{\pi }{4}}\frac{\Theta}{\Xi}\frac{c_k v_k}{t_k a_kd_k}, \quad g=\Theta\Xi\ .
\eeq
In the above, we have not restricted ourselves to the unitary case, hence the parameters $\eta_i$ are unconstrained. We remark that the parameters $\eta_i$ are to be identified with the $\eta(p_i)$ appearing in \cite{Arutyunov:2006yd}, even though here they are independent parameters and not functions of the momenta. With the identifications \eqref{eq:variablesidentifications}, the gluing condition \eqref{eq:gluingcond} becomes the mass shell condition:
\beq
\label{eq:massshellcondition}
x_k^++\frac{1}{x_k^+}-x_k^--\frac{1}{x_k^-}=\frac{i}{g}. 
\eeq

\subsection{Connections to the quantum symmetry}
\label{sec:connquantumsym}

In the previous sections, we established that the $\check{\tR}$ matrix of \eqref{eq:Rcheckfermionic} that we obtained from the Shastry-Shiroishi-Wadati construction of \eqref{eq:generalRmatrixfermionic} is invariant under the centrally extended superalgebra $\fh=$su$(2|2)\ltimes \mathbb{R}^2$. In fact, from \cite{Beisert:2006qh, Arutyunov:2006yd}, we know that it is uniquely determined by the symmetry requirement. On the other hand, in section \ref{subsec:quantumsymmetry} we found that the building blocks of the $\tR$ matrix are invariant under the affine quantum group $\hat{\fg}$, broken to just $\fg$ by the construction in \eqref{eq:Rcheckfermionic}. Now, we would like to carefully connect the two.  It turns out that we can relate all non-diagonal generators of $\hat{\fg}\oplus \hat{\fg}$ to the fermionic generators of su$(2|2)\ltimes \mathbb{R}^2$. Using \eqref{rep-fermi}, \eqref{eq:fermionicgenerators} as well as table \ref{table1}, we find 
\begin{align}
\label{eq:relQuantumpsl22}
&\cS^{1}_{\phantom{1}1}(\mathfrak{B})=\sqrt{2\Theta\Xi}\textsf{t}^{-1}\tf_{0,\uparrow}\textsf{t},& &\cS^{2}_{\phantom{2}1}(\mathfrak{B})=-\sqrt{2\Theta\Xi} \textsf{t}^{-1}\te_{0,\downarrow}\tk_{0,\downarrow}^{-1}\textsf{t},&\nonumber\\
&\cS^{1}_{\phantom{1}2}(\mathfrak{B})=\sqrt{2\Theta\Xi} \textsf{t}^{-1}\tf_{0,\downarrow}\textsf{t},& &\cS^{2}_{\phantom{2}2}(\mathfrak{B})=\sqrt{2\Theta\Xi} \textsf{t}^{-1}\te_{0,\uparrow}\tk_{0,\uparrow}^{-1}\textsf{t},&\nonumber\\
&\cQ^{1}_{\phantom{1}1}(\mathfrak{B})=\sqrt{\frac{2}{\Theta\Xi}} \textsf{t}^{-1}\textsf{x}^{-1}\tf_{1,\uparrow}\textsf{x}\textsf{t},& &\cQ^{1}_{\phantom{1}2}(\mathfrak{B})=\sqrt{\frac{2}{\Theta\Xi}} \textsf{t}^{-1}\textsf{y}^{-1}\te_{1,\downarrow}\tk_{1,\downarrow}^{-1}\textsf{y}\textsf{t},&\nonumber\\
&\cQ^{2}_{\phantom{2}1}(\mathfrak{B})=\sqrt{\frac{2}{\Theta\Xi}} \textsf{t}^{-1}\textsf{x}^{-1}\tf_{1,\downarrow}\textsf{x}\textsf{t},& &\cQ^{2}_{\phantom{2}2}(\mathfrak{B})=-\sqrt{\frac{2}{\Theta\Xi}} \textsf{t}^{-1}\textsf{y}^{-1}\te_{1,\uparrow}\tk_{1,\uparrow}^{-1}\textsf{y}\textsf{t},&
\end{align}
where the matrices $\mathfrak{B}$ were defined in \eqref{eq:defofthematricesBC}. Here, we have suppressed the lattice indices in order to improve readability. Furthermore, in \eqref{eq:relQuantumpsl22}, we introduced the operators $\textsf{t}_r\colonequals \tG_{r,\uparrow}\tG_{r,\downarrow} \tU_r$ as well as $\textsf{x}_r$ and $\textsf{y}_r$:
\beqa
\textsf{x}_r&\colonequals&\Theta^2\Xi^2x_r^{-2}\m_{r,\uparrow}\m_{r,\downarrow}+\Xi^2\lambda_rv_r^{-1}\m_{r,\uparrow}\n_{r,\downarrow}+\Xi^2\lambda_rv_r^{-1}\n_{r,\uparrow}\m_{r,\downarrow}+x_r^2\n_{r,\uparrow}\n_{r,\downarrow},\nonumber\\
\textsf{y}_r&\colonequals&x_r^{-2}\m_{r,\uparrow}\m_{r,\downarrow}+\Xi^2\lambda_rv_r^{-1}\m_{r,\uparrow}\n_{r,\downarrow}+\Xi^2\lambda_rv_r^{-1}\n_{r,\uparrow}\m_{r,\downarrow}+\Theta^2\Xi^2x_r^2\n_{r,\uparrow}\n_{r,\downarrow}.
\eeqa
Having thus established a direct connection between the algebras, we would like to relate the invariance conditions  \eqref{eq:remainingQuantumSymmetryR} and \eqref{eq:ferminvariance}. We described in section \ref{sec:invboscond} how to write the operator $\tR_{12}$ as a function of $\check{\tR}_{12}$ and inserting the result into \eqref{eq:remainingQuantumSymmetryR} leads to
\beq
\check{\tR}_{12}\left(\textsf{T}^{-1}\Delta(\tJ)\textsf{T}\right)=\left(\tV_2^{-1}\tP_{12}\tV_2\right)\left(\textsf{T}^{-1}\Delta^{\prime}(\tJ)\textsf{T}\right)\left(\tV_2^{-1}\tP_{12}\tV_2\right)\check{\tR}_{12}, \quad \forall \tJ\in\fg\oplus \fg,
\eeq
with $\textsf{T}:=\textsf{t}_1\textsf{t}_2\tV_2$. The appearance of the similarity transformation $\tV$ acting only on the second lattice site can be tracked back to the central operator $\textsf{Z}$ that we had to introduce in the quantum group coproduct back in \eqref{eq:Delta2}. We can now identify the coproducts of the quantum group elements to the left hand side of the invariance equation \eqref{eq:ferminvariance}. A straightforward computation using \eqref{rep-fermi}, \eqref{eq:Delta2}, and \eqref{eq:fermionicgenerators} leads to 
\begin{align}
\begin{split}
\big(\cS^{1}_{\phantom{1}1}\big)_1(\mathfrak{D}_1)+
\big(\cS^{1}_{\phantom{1}1}\big)_2(\mathfrak{D}_2)
&=-i\sqrt{2\Theta\Xi} 
\left(\textsf{T}^{-1}
\Delta(\tf_{0,\uparrow})\Delta(\tF_{\uparrow})\textsf{T}\right),\\
\big(\cS^{1}_{\phantom{1}2}\big)_1(\mathfrak{D}_1)+
\big(\cS^{1}_{\phantom{1}2}\big)_2(\mathfrak{D}_2)
&=-i\sqrt{2\Theta\Xi}
\left(\textsf{T}^{-1}
\Delta(\tf_{0,\downarrow})\Delta(\tF_{\downarrow})\textsf{T}\right),\\
\big(\cS^{2}_{\phantom{2}1}\big)_1(\mathfrak{D}_1)+
\big(\cS^{2}_{\phantom{2}1}\big)_2(\mathfrak{D}_2)
&=i\sqrt{2\Theta\Xi}
\left(\textsf{T}^{-1}
\Delta(\te_{0,\downarrow}\tk_{0,\downarrow}^{-1})
\Delta(\tF_{\downarrow})\textsf{T}\right),\\
\big(\cS^{2}_{\phantom{2}2}\big)_1(\mathfrak{D}_1)+
\big(\cS^{2}_{\phantom{2}2}\big)_2(\mathfrak{D}_2)
&=-i\sqrt{2\Theta\Xi}
\left(\textsf{T}^{-1}
\Delta(\te_{0,\uparrow} \tk_{0,\uparrow}^{-1})
\Delta(\tF_{\uparrow}) \textsf{T}\right), 
\end{split}
\label{actionSco}
\end{align}
where we have made the choice\footnote{We remind that the symmetry conditions \eqref{eq:symmetryconditions} imply $z^2=-1$ for the quantum group variables.} $z=i$ for the central element of the quantum group. Similar equations can also be written for the $\Delta^{\prime}$ factors. These formulas thus establish a direct link between the invariance of the Shiroishi and Wadati operator $\tR$ under the quantum group symmetry $\fg\oplus \fg$, \eqref{eq:remainingQuantumSymmetryR}, and the invariance of the AdS/CFT S-matrix under half of the $\fh$ fermionic generators, \eqref{eq:ferminvariance}.

We are then led to the question of whether there exists a similar connection between the remaining fermionic generators of $\fh$ and the broken symmetries of the affine quantum group $\hat{\fg}\oplus \hat{\fg}$. It turns out that, if we define $\textsf{X}:=\textsf{x}_1\textsf{x}_2$ and $\textsf{Y}:=\textsf{y}_1\textsf{y}_2$, we find for the remaining fermionic generators the relations:
\begin{align}
\begin{split}
\big(\cQ^{1}_{\phantom{1}1}\big)_1(\mathfrak{D}_1)+
\big(\cQ^{1}_{\phantom{1}1}\big)_2(\mathfrak{D}_2)
&=\sqrt{\frac{2}{\Theta\Xi}}
\left(\textsf{T}^{-1}\textsf{X}^{-1} 
\Delta(\tf_{1,\uparrow})
\Delta(\tF_{\uparrow})\textsf{X}\textsf{T}\right), 
\\
\big(\cQ^{2}_{\phantom{2}1}\big)_1(\mathfrak{D}_1)+
\big(\cQ^{2}_{\phantom{2}1}\big)_2(\mathfrak{D}_2)
&=\sqrt{\frac{2}{\Theta\Xi}}
\left(\textsf{T}^{-1}\textsf{X}^{-1}
 \Delta(\tf_{1,\downarrow})
\Delta(\tF_{\downarrow})\textsf{X}\textsf{T}\right),\\ 
\big(\cQ^{1}_{\phantom{1}2}\big)_1(\mathfrak{D}_1)+
\big(\cQ^{1}_{\phantom{1}2}\big)_2(\mathfrak{D}_2)
&=-\sqrt{\frac{2}{\Theta\Xi}}
\left(\textsf{T}^{-1}\textsf{Y}^{-1} 
\Delta(\te_{1,\downarrow} \tk_{1,\downarrow}^{-1})
\Delta( \tF_{\downarrow} )
\textsf{Y}\textsf{T}\right), 
\\
\big(\cQ^{2}_{\phantom{2}2}\big)_1(\mathfrak{D}_1)+
\big(\cQ^{2}_{\phantom{2}2}\big)_2(\mathfrak{D}_2)
&=\sqrt{\frac{2}{\Theta\Xi}}
\left(\textsf{T}^{-1}\textsf{Y}^{-1} 
\Delta(\te_{1,\uparrow} \tk_{1,\uparrow}^{-1})
\Delta( \tF_{\uparrow} )
\textsf{Y}\textsf{T}\right).
 \label{actionQco}
\end{split}
\end{align}
Hence, up to these similarity transformations, we can relate every generator of $\hat{\fg}\oplus \hat{\fg}$ to an odd element of $\fh$. To summarize, we find that the $\cS$ elements of $\fh$ are directly linked to the unbroken $\fg\oplus \fg$ generators, while the $\cQ$ ones become symmetries of the Shastry-Shiroishi-Wadati R-matrix only after an appropriate similarity transformation has been applied.

\section{The two layer structure in the AdS variables}

Our goal in this section is to provide a direct connection between the two-layer formulation and the one commonly used in the AdS/CFT literature. For that, we wish to rewrite the operator $\check{\tR}$ of \eqref{eq:Rcheckfermionic} using the variable identifications of \eqref{eq:variablesidentifications}.
After a rescaling, we identify
\footnote{See appendix \ref{connection-AFZ}}
 $\tP\check{\tR}$ with the matrix $S$ of \cite{Arutyunov:2006yd} and notice that $\check{\tR}$ depends on more parameters than $S$. Thus, we can set some of our parameters to special values without damaging the essence of the identification. A very symmetric choice is the following:
\beq
\Theta=\Xi=\sqrt{g} \quad \text{ and } \quad t_k=\sqrt{\frac{x_k^+}{\eta_k}}. 
\eeq
This allows us to invert \eqref{eq:variablesidentifications} and get
\beq
v_k=x_k^-, \quad a_k=\frac{\sqrt{i\eta_kx_k^-}}{x_k^--x_k^+}, \quad b_k=\sqrt{\frac{x_k^+}{i\eta_k}}, 
\quad c_k=\frac{\sqrt{i\eta_k x_{k}^+}}{x_k^--x_k^+}, 
\quad d_{k}=\sqrt{\frac{x_k^-}{i\eta_k}}.
\eeq
Plugging the above into the definition of the $\tR^{\pm}$ operators of \eqref{eq:defRpm}, we get for each layer 
\beqa
\tR_{12}^{+}&=&-\sqrt{\frac{\eta_2}{\eta_1}}\frac{1}{x_2^--x_2^+}\left(\sqrt{x_2^-}\n_1+i \sqrt{x_2^+}\m_1\right)\left(\sqrt{x_1^-}\n_2-i \sqrt{x_1^+}\m_2\right)+\c_1^{\dagger}\c_2,\nonumber\\
\tR_{12}^{-}&=&\sqrt{\frac{\eta_1}{\eta_2}}\frac{1}{x_1^--x_1^+}\left(\sqrt{x_2^+}\n_1+i \sqrt{x_2^-}\m_1\right)\left(\sqrt{x_1^+}\n_2-i \sqrt{x_1^-}\m_2\right)+\c_2^{\dagger}\c_1\ .
\eeqa
Furthermore, the similarity transformation matrices of \eqref{eq:simtranfUV} become
\beqa
\tU_k&=& \m_{k,\uparrow}\m_{k,\downarrow}+\sqrt{\frac{x_k^+}{\eta_k}}\left( \m_{k,\uparrow}\n_{k,\downarrow}+ 
\n_{k,\uparrow}\m_{k,\downarrow}\right)+
 \frac{i \eta_k}{x_k^+-x_k^-}\n_{k,\uparrow}\n_{k,\downarrow}
\nonumber\\
\tV_k&=& ( \m_{k,\uparrow}-i  \n_{k,\uparrow})( \m_{k,\downarrow}-i  \n_{k,\downarrow})\ .
\eeqa
After expressing the operator  $\check{\tR}_{12}$ in the new variables, 
we divide it by the factor 
$\frac{x_1^+x_2^+-x_1^-x_2^-}{(x_1^--x_1^+)(x_2^--x_2^+)}
\frac{x_1^- - x_2^+}{x_1^-+x_2^+}$ 
and denote it again as $\check{\tR}_{12}$. 
The final expression then reads
\beqa
\label{eq:finalexpressionSmatrix}
\check{\tR}_{12}&=
&\frac{(x_1^--x_1^+)(x_2^--x_2^+)}{x_1^+x_2^+-x_1^-x_2^-}\tV_2^{-1}\tP_{12} \tU_1^{-1}\tU_2^{-1}\left\{\frac{x_2^-+x_1^+}{x_1^--x_2^+}\left(\frac{x_2^+\eta_1}{x_1^+\eta_2}\frac{e^{i\frac{p_2}{2}}-e^{-i\frac{p_2}{2}}}{e^{i\frac{p_1}{2}}-e^{-i\frac{p_1}{2}}}\tR^+_{12,\uparrow}\tR^+_{12,\downarrow}\right.\right.\nonumber\\
&&\left.\left.
+\frac{x_1^- \eta_2}{x_2^- \eta_1}
\frac{e^{i\frac{p_1}{2}}-e^{-i\frac{p_1}{2}}}
{e^{i\frac{p_2}{2}}-e^{-i\frac{p_2}{2}}}\tR^-_{12,\uparrow}\tR^-_{12,\downarrow}\right)+\frac{x_2^++x_1^-}{x_1^--x_2^+}\left(\tR^+_{12,\uparrow}\tR^-_{12,\downarrow}+\tR^-_{12,\uparrow}\tR^+_{12,\downarrow}\right)\right\}\tU_1\tU_2\tV_2\ ,\nonumber\\
\eeqa
where as usual the worldsheet momentum is given by the rapidity variables as 
\begin{align}
e^{ip_{k}}=\frac{x^+_k}{x^-_k}, \qquad k=1,2 .
 \label{WS-momentum}
\end{align}
Thus, the above equation \eqref{eq:finalexpressionSmatrix} finally makes completely explicit the two-layer structure of the AdS/CFT S-matrix.

\section{Outlook}
In this paper we have (re)constructed the S-matrix for $AdS_{5} \times S^{5}$ without relying on the central extension of $su(2|2)$ \cite{Beisert:2006qh}. 
It turned out that the AdS/CFT S-matrix \cite{Beisert:2005tm,Arutyunov:2006yd} is essentially a special case of Shiroishi and Wadati's generalized Hubbard R-matrix \cite{Shiroishi1995} which appeared about 10 years earlier. We had to impose the symmetry condition \eqref{eq:symmetryconditions} on Shiroishi and Wadati's R-matrix to obtain the AdS/CFT S-matrix. The only known S-matrix\footnote{that intertwines between two four-dimensional vector spaces.} which contains the AdS/CFT S-matrix other than Shiroishi and Wadati's R-matrix is the q-deformed S-matrix proposed in \cite{Beisert:2008tw}. In this context, whether relaxing the symmetry condition corresponds to the q-deformation or not is an interesting open question. 

An ambitious goal will be to construct an infinite-dimensional R-matrix for the AdS/CFT correspondence in a multilayer approach, possibly in a four layer model, by generalizing our formalism. This R-matrix would of course need to realize all the intrinsic structures suggested by the asymptotic Bethe Ansatz equations \cite{Beisert:2005fw}, the Y-system \cite{GKV09}, thermodynamic Bethe Ansatz equations \cite{BFT09}, nonlinear integral equations \cite{Gromov:2011cx} and a  group theoretical argument on characters \cite{Gromov:2010vb}. Some help in that direction might come by a better understanding of the role that the tetrahedron Zamolodchikov equations, as well as the three-dimensional integrability structures they allude to, play in the AdS/CFT correspondence. Bazhanov and Sergeev obtained in \cite{Bazhanov:2005as} solutions of the tetrahedron equations systematically and uncovered a relation to the quantum affine algebra $\text{U}_{q}(\widehat{\text{sl}}(n))$. It will be very interesting to see how their method fits into our problem.

\section*{Acknowledgments}
The authors thank Vladimir  Bazhanov, Vladimir Kazakov and Anton Zabrodin for discussions. 
We also thank Takuya Matsumoto for a correspondence on \cite{Beisert:2011wq}. 
The work of ZT is supported by SFB 647 ``Space-Time-Matter''. The research leading to these results has received funding
\footnote{The text was revised after ZT moved to 
Department of Theoretical Physics, RSPE, 
Australian National University 
(where he was supported by the Australian Research Council),  
School of Mathematics and Statistics, 
The University of Melbourne 
(where he was supported by the Australian Research Council), 
Fakult\"at f\"ur Mathematik und Naturwissenschaften, 
Bergische Universit\"at Wuppertal (where he was supported by the university), and 
Laboratoire de Math\'ematiques et Physique Th\'eorique CNRS/UMR 7350,
 F\'ed\'eration Denis Poisson FR2964,
Universit\'e de Tours (where he was supported by CNRS).  
He thanks V.Bazhanov, O.Foda, H.Boos and P.Baseilhac  
for giving him opportunities to work at these institutions. 
The research of Z.T. is supported by the European Research Council
(Programme �gIdeas�h ERC-2012-AdG 320769 AdS-CFT-solvable).} 
from the European Union Marie Curie International Research Staff Exchange Scheme UNIFY  under grant agreement n$^{\rm o}$~269217.

\appendix
\section{Jordan-Wigner transformation}
\label{Jordan}
\addcontentsline{toc}{section}{Appendix A}
\def\theequation{A\arabic{equation}}
\setcounter{equation}{0}
Unlike in the rest of the main text, here we prefer to work with ordinary matrices instead of oscillators. To that end, we introduce the so-called Jordan-Wigner transformation 
(see for example, an appendix 
in \cite{Umeno1998} which is relevant to our discussion),
\begin{align}
\begin{split}
\c_{j}&=
   \left( \prod_{k=1}^{j-1}  \sigma^{z}_{k} \right)
 \sigma^{-}_{j}  , 
\qquad 
\c^{\dagger}_{j}=
 \left( \prod_{k=1}^{j-1}\sigma^{z}_{k} \right) \sigma^{+}_{j} ,
\end{split}
\label{JW-trans0}
\end{align}
where $\sigma^{z}_{j}$ and $\sigma^{\pm}_{j}$ are $2 \times 2$ 
matrices 
$\sigma^{z}=\begin{pmatrix}
1 & 0  \\
0 & -1  \\
\end{pmatrix}$, 
$\sigma^{+}=\begin{pmatrix}
0 & 1  \\
0 & 0  \\
\end{pmatrix}$ and 
$\sigma^{-}=\begin{pmatrix}
0 & 0  \\
1 & 0  \\
\end{pmatrix}$ at the lattice site $j \in \{1,2,\dots, N \}$. 
Note that 
\eqref{JW-trans0} realizes the relation \eqref{eq:formionicoscillators}. 
In order to apply the transformation \eqref{JW-trans0} to
the operator ${\mathsf R}^{\text{f}}_{12}(A)$ \eqref{fermionicRmatrix}, 
one has to multiply the graded permutation \eqref{gr-permutation} first. 
After applying \eqref{JW-trans0}, one also has to multiply the non-graded permutation 
$P=\sum_{i,j=1}^{2}e_{ij} \otimes e_{ji}$, where $e_{ij}$ is a $2 \times 2$ matrix unit 
whose $(k,l)$ matrix element is $\delta_{ik}\delta_{jl}$. 
Then the operator ${\mathsf R}^{\text{f}}_{12}(A)$ in \eqref{fermionicRmatrix} is mapped
\footnote{This is for $N=2$ case; 
or a factor $ \mathbbm{1}_{2}^{\otimes (N-2)} $ is omitted for $N>2$, 
where $ \mathbbm{1}_{2} := e_{11}+e_{22}$.} 
to the R-matrix of the free fermion 6-vertex model: 
\begin{align}
R(A) :=
P\left[ {\mathsf P}_{12}{\mathsf R}^{\text{f}}_{12}(A)\right]_{\eqref{JW-trans0}}=
\begin{pmatrix}
a & 0 & 0 & 0 \\
0 & ib & 1 & 0 \\
0 & 1 & ic & 0 \\
0 & 0 & 0 & d
\end{pmatrix}
, 
\qquad ad-bc=1.
\label{R-6vertex}
\end{align}
Then we can introduce the matrices 
$R^{0}(A_{1},A_{2}):=R(A_{2}A_{1}^{-1}) $ as well as 
$R^{1}(A_{1},A_{2}):=R(A_{2}\sigma^{z}A_{1}^{-1}\sigma^{z})(1 \otimes \sigma^{z}) $ 
for 
$A_{j}=
\begin{pmatrix}
a_{j} & b_{j}  \\
c_{j} & d_{j}  \\
\end{pmatrix}
\in \text{SL}(2,{\mathbb C})$, 
which correspond to their oscillator counterparts of \eqref{r0r1}. 
We will also use  
$ R^{\pm }(A_{1},A_{2}) =\frac{1}{2}(R^{0}(A_{1},A_{2}) \pm R^{1}(A_{1},A_{2}) ) $, 
which are given by 
\begin{align}
R^{+}(A_{1},A_{2})&=\begin{pmatrix}
a_{2}d_{1} & 0 & 0 & 0 \\
0 & -ia_{2}b_{1} & 1 & 0 \\
0 & 0 & ic_{2}d_{1} & 0 \\
0 & 0 & 0 & -c_{2}b_{1}
\end{pmatrix}
, \label{Rp-spin}
\\[6pt]
R^{-}(A_{1},A_{2})&=\begin{pmatrix}
-b_{2}c_{1} & 0 & 0 & 0 \\
0 & ib_{2}a_{1} & 0 & 0 \\
0 & 1 & -id_{2}c_{1} & 0 \\
0 & 0 & 0 & d_{2}a_{1}
\end{pmatrix}
. 
\label{Rm-spin}
\end{align}  
These R-matrices \eqref{Rp-spin} and \eqref{Rm-spin} 
are related to \eqref{R-6vertex} as
$ R(A)=R^{+}(1,A)+R^{-}(1,A)$. 

The Jordan-Wigner transformation 
\eqref{JW-trans0} can be generalized to a two-layer lattice, 
\begin{align}
\begin{split}
\c_{j,\uparrow}&=
   \left( \prod_{k=1}^{j-1}  \sigma^{z}_{k} \right)
 \sigma^{-}_{j} \otimes 1 , 
\qquad 
\c^{\dagger}_{j,\uparrow}=
 \left( \prod_{k=1}^{j-1}\sigma^{z}_{k} \right) \sigma^{+}_{j} \otimes 1,
\\
\c_{j,\downarrow} &= 
\left( \prod_{k=1}^{N}\sigma^{z}_{k}  \right) \otimes  
\left( \prod_{k=1}^{j-1}\sigma^{z}_{k} \right) \sigma^{-}_{j} , 
\qquad
\c^{\dagger}_{j,\downarrow} =
\left( \prod_{k=1}^{N}\sigma^{z}_{k}   \right) \otimes  
\left( \prod_{k=1}^{j-1}\sigma^{z}_{k} \right) \sigma^{+}_{j}. 
\end{split}
\label{JW-trans}
\end{align}
Note that \eqref{JW-trans} realizes  the relations  \eqref{eq:formionicoscillators2}. \eqref{JW-trans} are matrix representation
\footnote{Here we define the matrix expression of any operator ${\mathbf A} $ 
on the Fock space as 
${\mathbf A} (\c^{\dagger} \left|0 \right \rangle , \left|0 \right \rangle ) = 
 ( \c^{\dagger} \left|0 \right \rangle , \left|0 \right \rangle ) 
 \begin{pmatrix} 
 a_{11} & a_{12} \\
 a_{21} & a_{22} 
 \end{pmatrix} $ 
 on each lattice site.} 
of 
the fermion operators on the basis 
\begin{align}
\left|n_{1,\uparrow},\dots, n_{N,\uparrow}, 
n_{1,\downarrow}, \dots, n_{N,\downarrow} \right \rangle 
=(\c^{\dagger}_{1,\uparrow})^{n_{1,\uparrow}}
\cdots
(\c^{\dagger}_{N,\uparrow})^{n_{N,\uparrow}}
(\c^{\dagger}_{1,\downarrow})^{n_{1,\downarrow}} 
\cdots
(\c^{\dagger}_{N,\downarrow})^{n_{N,\downarrow}} 
\left|0 \right \rangle 
\end{align}
in the Fock space, where 
$n_{j,\uparrow},n_{j,\downarrow} \in \{0,1 \}$, 
$\c_{j,\uparrow} \left|0 \right \rangle =\c_{j,\downarrow} \left|0 \right \rangle =0 $.

\section{On the graded Yang-Baxter equation}
\label{cone-gr-YBE}
\addcontentsline{toc}{section}{Appendix B}
\def\theequation{B\arabic{equation}}
\setcounter{equation}{0}
In this appendix, we briefly summarize the 
relation between the graded Yang-Baxter equation 
and the non-graded Yang-Baxter equation (cf. \cite{KuSk82}). 

Let  $N$ be a non-negative integer. 
 We define  
 the grading parameter $p$ on the set  $\{1,2,\dots N \}$, namely 
 for any $j \in \{1,2,\dots N \}$, we set $p(j)=0$ or $p(j)=1$.
Let $E_{ij}$ be $N \times N $ matrix units
 and $e_i$ be canonical basis vectors\footnote{The discussion in this appendix is valid for any non-negative integer $N$. 
However, we will use the notations $p,E_{ij}$ and $e_{j}$ only for $N=4$ in the subsequent sections.}
${\mathbb C}^{N}$, 
 which satisfy $E_{ij}e_{k}=\delta_{jk}e_{i}$. 
Then we define the grading of these object by  
$p(e_{j})=p(j)$ and  $p(E_{ij})=p(i) +p(j) \mod 2$. 
 For any homogeneous elements\footnote{We call an object whose grading can be uniquely defined homogeneous.} 
 $B_{i}$, 
the graded (super) tensor product is defined by 
 $(B_{1} \otimes_{s} B_{2}) (B_{3} \otimes_{s} B_{4}) =(-1)^{p(B_{2})p(B_{3}) }(B_{1} B_{3} \otimes_{s} B_{2}B_{4}) $.  We define the matrix elements 
$\left[ {\mathbf R}^{g}_{12}\right]^{ i_{1},i_{2} }_{j_{1},j_{2} }$ of an R-matrix $ {\bf R}^{g}_{12} \in  
 \mathrm{End}({\mathbb C}^{N} \otimes_{s} {\mathbb C}^{N})$ by 
\begin{align}
 {\bf R}^{g}_{12} (e_{j_{1}} \otimes_{s} e_{j_{2}}) 
 =\sum_{i_{1},i_{2} =1}^{N} (e_{i_{1}} \otimes_{s} e_{i_{2}})  
 \left[ {\mathbf R}^{g}_{12}\right]^{ i_{1},i_{2} }_{j_{1},j_{2} } .
\end{align}
Note that 
${\mathbf R}^{g}_{12}$ is realized by 
 $ {\mathbf R}^{g}_{12}=
 \sum_{i_{1},i_{2},j_{1},j_{2}}
 (-1)^{p(j_{1})(p(i_{2}) +p(j_{2}) ) }
  \left[ {\mathbf R}^{g}_{12}\right]^{i_{1}i_{2}}_{j_{1}j_{2}} \ 
  E_{i_{1}j_{1}} \otimes_{s}  E_{i_{2}j_{2}} $ 
 on the ${\mathbb Z}_{2}$-graded vector space. 
 Suppose that the R-matrix ${\mathbf R}^{g}_{12} $ 
 satisfies\footnote{Here we abuse notations so that the subscripts $i,j$ in $ {\mathbf R}^{g}_{ij}$   denote both the space and the spectral parameter dependence of the R-matrix. Thus we have to interpret the R-matrices as $ {\mathbf R}^{g}_{12}=
 \sum_{i_{1},i_{2},j_{1},j_{2}}
 (-1)^{p(j_{1})(p(i_{2}) +p(j_{2}) ) }
  \left[ {\mathbf R}^{g}_{12}\right]^{i_{1}i_{2}}_{j_{1}j_{2}} \ 
  E_{i_{1}j_{1}} \otimes_{s}  E_{i_{2}j_{2}} \otimes_{s} \mathbbm{1}$ as well as 
 $ {\mathbf R}^{g}_{13}=
 \sum_{i_{1},i_{3},j_{1},j_{3}}
 (-1)^{p(j_{1})(p(i_{3}) +p(j_{3}) ) }
  \left[ {\mathbf R}^{g}_{13}\right]^{i_{1}i_{3}}_{j_{1}j_{3}} \ 
  E_{i_{1}j_{1}} \otimes_{s} \mathbbm{1} \otimes_{s}  E_{i_{3}j_{3}} $ and 
 $ {\mathbf R}^{g}_{23}=
 \sum_{i_{2},i_{3},j_{2},j_{3}}
 (-1)^{p(j_{2})(p(i_{3}) +p(j_{3}) ) }
  \left[ {\mathbf R}^{g}_{23}\right]^{i_{2}i_{3}}_{j_{2}j_{3}} \ 
 \mathbbm{1} \otimes_{s}  E_{i_{2}j_{2}} \otimes_{s}  E_{i_{3}j_{3}} $. 
 The same type of convention will be used for the Yang-Baxter equation 
 in what follows. 
  }
the graded Yang-Baxter equation 
 $
 {\mathbf R}^{g}_{12}{\mathbf R}^{g}_{13}{\mathbf R}^{g}_{23}=
 {\mathbf R}^{g}_{23}{\mathbf R}^{g}_{13}{\mathbf R}^{g}_{12}
 $. In components, it can be rewritten as
\begin{multline}
\sum_{j_{1},j_{2},j_{3} }
\left[ {\mathbf R}^{g}_{12}\right]^{ i_{1},i_{2} }_{j_{1},j_{2} }
  \left[ {\mathbf R}^{g}_{13}\right]^{ j_{1},i_{3} }_{k_{1},j_{3} }
   \left[ {\mathbf R}^{g}_{23}\right]^{ j_{2},j_{3} }_{k_{2},k_{3} } 
   (-1)^{p(j_{2})(p(i_{3}) +p(j_{3})) +p(k_{1}) (p(j_{2}) +p(k_{2}) +p(j_{3})+p(k_{3}) )} 
   \\
   =
\sum_{j_{1},j_{2},j_{3} }
\left[ {\mathbf R}^{g}_{23}\right]^{ i_{2},i_{3} }_{j_{2},j_{3} }
  \left[ {\mathbf R}^{g}_{13}\right]^{ i_{1},j_{3} }_{j_{1},k_{3} }
   \left[ {\mathbf R}^{g}_{12}\right]^{ j_{1},j_{2} }_{k_{1},k_{2} } 
   (-1)^{p(j_{2})(p(j_{3}) +p(k_{3})) +p(i_{1}) (p(i_{2}) +p(j_{2}) +p(i_{3})+p(j_{3}) )} . 
    \label{gr-YBE-com}
\end{multline}
 This relation \eqref{gr-YBE-com} reduces to a more familiar form 
\begin{multline}
\sum_{j_{1},j_{2},j_{3} }
\left[ {\mathbf R}^{g}_{12}\right]^{ i_{1},i_{2} }_{j_{1},j_{2} }
  \left[ {\mathbf R}^{g}_{13}\right]^{ j_{1},i_{3} }_{k_{1},j_{3} }
   \left[ {\mathbf R}^{g}_{23}\right]^{ j_{2},j_{3} }_{k_{2},k_{3} } 
   (-1)^{p(j_{2})(p(i_{3}) +p(j_{3})) } 
   \\
   =
\sum_{j_{1},j_{2},j_{3} }
\left[ {\mathbf R}^{g}_{23}\right]^{ i_{2},i_{3} }_{j_{2},j_{3} }
  \left[ {\mathbf R}^{g}_{13}\right]^{ i_{1},j_{3} }_{j_{1},k_{3} }
   \left[ {\mathbf R}^{g}_{12}\right]^{ j_{1},j_{2} }_{k_{1},k_{2} } 
   (-1)^{p(j_{2})(p(j_{3}) +p(k_{3})) } 
    \label{gr-YBE-com2}
\end{multline}
under 
 the condition 
\begin{align}
p(i_{1}) + p(i_{2})+ p(j_{1})+ p(j_{2})=0 \mod 2 , 
\quad \text{for any} \quad  \left[ {\mathbf R}^{g}_{12}\right]^{ i_{1},i_{2} }_{j_{1},j_{2} } .
 \label{evenR}
\end{align} 
Let 
$\left[ {\mathbf R}_{12}\right]^{ i_{1},i_{2} }_{j_{1},j_{2} }=(-1)^{p(i_{1})p(i_{2}) }
\left[ {\mathbf R}^{g}_{12}\right]^{ i_{1},i_{2} }_{j_{1},j_{2} } $. 
It is well known that \eqref{gr-YBE-com2} (and \eqref{gr-YBE-com}) 
is equivalent to 
\begin{align}
\sum_{j_{1},j_{2},j_{3} }
\left[ {\mathbf R}_{12}\right]^{ i_{1},i_{2} }_{j_{1},j_{2} }
  \left[ {\mathbf R}_{13}\right]^{ j_{1},i_{3} }_{k_{1},j_{3} }
   \left[ {\mathbf R}_{23}\right]^{ j_{2},j_{3} }_{k_{2},k_{3} } 
   =
\sum_{j_{1},j_{2},j_{3} }
\left[ {\mathbf R}_{23}\right]^{ i_{2},i_{3} }_{j_{2},j_{3} }
  \left[ {\mathbf R}_{13}\right]^{ i_{1},j_{3} }_{j_{1},k_{3} }
   \left[ {\mathbf R}_{12}\right]^{ j_{1},j_{2} }_{k_{1},k_{2} } 
   \label{YBE-com}
\end{align}
under the condition \eqref{evenR}, and thus the R-matrix 
 $ {\mathbf R}_{12}=
 \sum_{i_{1},i_{2},j_{1},j_{2}}
  \left[ {\mathbf R}_{12}\right]^{i_{1}i_{2}}_{j_{1}j_{2}} \ 
  E_{i_{1}j_{1}} \otimes  E_{i_{2}j_{2}} \in \mathrm{End}({\mathbb C}^{N} \otimes {\mathbb C}^{N})$
 satisfies the non-graded Yang-Baxter equation 
 $
 {\mathbf R}_{12}{\mathbf R}_{13}{\mathbf R}_{23}=
 {\mathbf R}_{23}{\mathbf R}_{13}{\mathbf R}_{12}
 $.

\section{Connections to previous works}
\label{connection}
\addcontentsline{toc}{section}{Appendix C}
\def\theequation{C\arabic{equation}}
\setcounter{equation}{0}
In this appendix, we 
establish exact connections 
between the different notations used our main text and 
in  relevant previous works. 
\subsection{Shiroishi-Wadati \cite{Shiroishi1995}}
\label{connection-SW}
Let us multiply the fermionic R-matrix 
\eqref{eq:generalRmatrixfermionic} from the left by
the two-layer permutation operator
 $\tP_{12}= \tP_{12,\uparrow} \tP_{12,\downarrow}$
and apply the Jordan-Wigner transformation for 
$\tP_{12} \tR_{12}$. 
Then, multiplying the result from the left by the tensor product of a non-graded permutation $P \otimes P $ 
($P=\sum_{i,j=1}^{2} e_{ij} \otimes e_{ji}$; $e_{ij}$: $2\times 2$ matrix unit), we get the R-matrix of a spin model.
\begin{multline}
\label{eq:generalRmatrixSpin}
{\mathfrak R}_{12}:=(P \otimes P)
 \left[
\tP_{12} \tR_{12} 
\right]_{\eqref{JW-trans}} =
\\
=
\frac{v_2+\frac{b_1 c_1}{a_1 d_1} v_1}{\frac{b_2 c_2}{a_2 d_2} v_2+v_1}
\left(\frac{a_1 b_2}{b_1 a_2} 
 R^+_{12} \otimes R^+_{12} + 
 \frac{d_1 c_2}{c_1 d_2} 
 R^-_{12} \otimes  R^-_{12}\right) + 
  R^+_{12} \otimes  R^-_{12} + 
  R^-_{12} \otimes  R^+_{12}\ ,
\end{multline}
where $ R^{\pm }_{12}=R^{\pm}(A_{1},A_{2})$ are defined in \eqref{Rp-spin} and \eqref{Rm-spin}. 
We remark that this R-matrix satisfies the non-graded Yang-Baxter equation 
${\mathfrak R}_{12}{\mathfrak R}_{13}{\mathfrak R}_{23}
={\mathfrak R}_{23}{\mathfrak R}_{13}{\mathfrak R}_{12}$ 
under the gluing conditions \eqref{eq:gluingcond}.
The tetrahedral Zamolodchikov algebra for the R-matrices 
\eqref{Rp-spin} and \eqref{Rm-spin} 
has exactly the same form as 
\eqref{eq:tetrahedronZamalgebra}: 
\beq
\label{eq:tetrahedronZamalgebra-spin}
R^{\alpha}_{23} R^{\beta}_{13} R^{\gamma}_{12}=
\sum_{\bar{\alpha},\bar{\beta},\bar{\gamma}=\pm} 
\bS^{\alpha\beta\gamma}_{\bar{\alpha}\bar{\beta}\bar{\gamma}}\, 
R^{\bar{\gamma}}_{12} R^{\bar{\beta}}_{13} R^{\bar{\alpha}}_{23}\ ,
\eeq
where the coefficients $\bS^{\alpha\beta\gamma}_{\bar{\alpha}\bar{\beta}\bar{\gamma}}$ are given by \eqref{eq:defcoeffS}. 

The above R-matrix \eqref{eq:generalRmatrixSpin} 
 coincides with Shiroishi and Wadati's generalized R-matrix 
[eq.\ (4.28) in \cite{Shiroishi1995}] 
under the identification 
\begin{align}
\begin{split}
a^{+}_{j}&=a_{j}, 
\qquad 
a^{-}_{j}=d_{j},
\qquad 
b^{+}_{j}=e^{\frac{\pi i}{2}} b_{j},
\qquad 
b^{-}_{j}=e^{\frac{\pi i}{2}}  c_{j},
\\
t_{j}&= \sqrt{e^{i \pi} \frac{b_{j}c_{j}}{a_{j}d_{j}}} \, v_{j}, 
\qquad 
A=e^{\frac{\pi i}{2}}  \Theta^{2}, 
\qquad 
B=e^{\frac{\pi i}{2}}  \Xi^{2}, 
\qquad 
j=1,2,
\end{split}
\label{ident-SW}
\end{align}
where $t_{j}, a^{\pm}_{j}, b^{\pm}_{j}, A$ and $B$ are notations in \cite{Shiroishi1995}. 
Note that 
the gluing conditions \eqref{eq:gluingcond} and  
 the R-matrices 
\eqref{Rp-spin} and \eqref{Rm-spin} coincide 
with equations \ (4.27), (2.1) and (2.2) 
in \cite{Shiroishi1995}, respectively, 
under \eqref{ident-SW}. 
The coefficients 
$\bS^{\alpha\beta\gamma}_{\bar{\alpha}\bar{\beta}\bar{\gamma}}$ \eqref{eq:defcoeffS} 
of the tetrahedral Zamolodchikov algebra coincide with the coefficients 
$ S^{\alpha\beta\gamma}_{\bar{\alpha}\bar{\beta}\bar{\gamma}}$ 
 defined by eqs.\ (2.4), (2.13) and (2.15) in \cite{Shiroishi1995} 
 if we apply the cyclic shift $(1,2,3) \to (2,3,1)$ to the indices in \cite{Shiroishi1995}, 
 and consequently \eqref{eq:tetrahedronZamalgebra-spin} 
 coincides with the tetrahedral Zamolodchikov algebra written below 
 eq.\ (2.14) in \cite{Shiroishi1995}.

\subsection{Shastry \cite{Shastry1986} }
\label{connection-Sha}
Let us consider the limit \eqref{Hu-limit2}-\eqref{Hu-limit3} for \eqref{eq:generalRmatrixSpin} 
[cf.\ eq.\ (4.32) in \cite{Shiroishi1995}].
\begin{multline}
R^{\text{Sha}}_{12} :=
{\mathfrak R}_{12} |_{\eqref{Hu-limit2}, \eqref{Hu-limit3}}=
\\[6pt]
=
\overline{\rho}_{12} \left\{ 
\cosh (h_{2} -h_{1})\cos (u_{2} +u_{1}) \overline{R}^{0}_{12} \otimes \overline{R}^{0}_{12} 
 + \sinh (h_{2} -h_{1})\cos (u_{2} -u_{1}) \overline{R}^{1}_{12} \otimes \overline{R}^{1}_{12} 
 \right\}, 
 \\[6pt]
 \overline{\rho}_{12} =\frac{(e^{2h_{1}} +e^{2h_{2}} )(1- \tan u_{1} \tan u_{2})}
 { (e^{2h_{1}} - e^{2h_{2}} \tan u_{1} \tan u_{2})\cosh (h_{2}-h_{1} ) \cos (u_{2}+u_{1}) } ,
 \label{R-mat-Sha}
\end{multline}
where the R-matrices of the symmetric free fermion model are defined by 
\begin{align}
\overline{R}^{0}_{12}:& =
R^{0}_{12}(A_{1},A_{2}) |_{\eqref{Hu-limit2}} 
\nonumber \\[5pt]
&=
\begin{pmatrix}
\cos(u_{2}-u_{1}) & 0 & 0 & 0 \\
0 & \sin(u_{2}-u_{1}) & 1 & 0 \\
0 & 1 & \sin(u_{2}-u_{1}) & 0 \\
0 & 0 & 0 & \cos(u_{2}-u_{1})
\end{pmatrix}
,
\\[6pt]
\overline{R}^{1}_{12}:& =
R^{1}_{12}(A_{1},A_{2}) |_{\eqref{Hu-limit2}} 
\nonumber \\[5pt]
&=
\begin{pmatrix}
\cos(u_{2}+u_{1}) & 0 & 0 & 0 \\
0 &- \sin(u_{2}+u_{1}) & 1 & 0 \\
0 & -1 & \sin(u_{2}+u_{1}) & 0 \\
0 & 0 & 0 & -\cos(u_{2}+u_{1})
\end{pmatrix}
.
\end{align}
This R-matrix \eqref{R-mat-Sha} coincides with Shastry's R-matrix\footnote{Shastry's R-matrix in \cite{Shastry:1986zz} is related to the one in \cite{Shastry1986} 
by a simple gauge transformation mentioned below eq. (4.15) in \cite{Shastry1986}.}
(S-matrix)  [eq.\ (4.15) in \cite{Shastry1986}]  
under the identification 
\begin{multline}
\frac{R^{\text{Sha}}_{12}}{ \overline{\rho}_{12} } = 
\frac{S_{12}(\theta_{2}| \theta_{1}) }{ \rho }, 
\qquad 
\overline{R}^{0}_{12} \otimes \overline{R}^{0}_{12} =l_{12}(\theta_{2}-\theta_{1}), 
\qquad 
\overline{R}^{1}_{12} \otimes \overline{R}^{1}_{12} =l_{12}(\theta_{2}+\theta_{1})
  \sigma^{z}_{2} \tau^{z}_{2}, 
\\
 u_{j}=\theta_{j} ,  \quad j=1,2, 
\end{multline}
where the right hand side of these equations are written in terms of the notations in 
\cite{Shastry1986}. 
In addition, the gluing condition \eqref{Hu-limit4} coincides with eq.\ (4.14) in \cite{Shastry1986}. 

\subsection{Arutyunov-Frolov-Zamaklar \cite{Arutyunov:2006yd}}
\label{connection-AFZ}
Next we would like to mention a relation between our two-layer S-matrix $\check{\tR}_{12}$ 
\eqref{eq:finalexpressionSmatrix} and a S-matrix in \cite{Arutyunov:2006yd}. 
Let $S_{12}(p_{1},p_{2})$ be the S-matrix defined in eq.\ (A.7) [or eq.\ (8.7) in 
the arXiv version] in 
\cite{Arutyunov:2006yd}. We adopt the `string basis' (in their terminology) and 
assume that the relation\footnote{The notation $\eta(p_{k})$ in \cite{Arutyunov:2006yd} is 
denoted as $\eta_{k} $ in this paper.
}
\begin{align}
\eta_{k}=\sqrt{i x_{k}^{-} -i x_{k}^{+} } 
 \label{AFZ-unitary}
\end{align}
follows from unitary representations. 
Let $E_{ij}$ be $4 \times 4$ matrix unit. 
Then we introduce the following matrices 
\begin{align}
\bar{P}_{12}&=
\mathbbm{1}_{2} 
\otimes P \otimes
\mathbbm{1}_{2} , 
 \label{Pbar}
\\[6pt]
{\mathcal C} &= E_{11} +E_{42} +E_{33}  +E_{24}  
, \label{shuffle}
\\[6pt]
{\mathcal F}_{12} &=\sum_{j,k=1}^{4} \varepsilon_{jk} E_{jj} \otimes E_{kk} , 
\qquad 
{\mathcal F}_{21} =\sum_{j,k=1}^{4} \varepsilon_{kj} E_{jj} \otimes E_{kk} , 
\label{signtwist}
\end{align}
where\footnote{There is a freedom for the choices of these matrices. 
One can replace the matrix \eqref{shuffle} with 
${\mathcal C} = E_{11} +E_{32} +E_{43}  +E_{24} $. 
Moreover, one can also choose 
${\mathcal C} = E_{21} +E_{32} +E_{43}  +E_{14} $ or 
${\mathcal C} = E_{21} +E_{42} +E_{33}  +E_{14} $ 
if the sign function for  \eqref{signtwist} 
is replaced by $\varepsilon_{jk}=-1$  for $ (j,k)=(3,1),(4,1),(4,3)$, otherwise  
$\varepsilon_{jk}=1$.
}
 $\varepsilon_{jk}=-1$ for $ (j,k)=(3,2),(3,4),(4,2)$, otherwise  
$\varepsilon_{jk}=1$.
We find that the following relation holds between these two S-matrices 
under the relations \eqref{eq:massshellcondition} and \eqref{WS-momentum}.
\begin{align}
S_{12}(p_{1},p_{2})=
{\mathcal F}_{21}
({\mathcal C} \otimes {\mathcal C})
\bar{P}_{12}
(P \otimes P)
 \left[ \check{\tR}_{12}
\right]_{\eqref{JW-trans}}
\bar{P}_{12}
({\mathcal C}^{-1} \otimes {\mathcal C}^{-1} )
{\mathcal F}_{12}^{-1} .
\label{eq:AFZ-Ours}
\end{align}
On the right hand side of \eqref{eq:AFZ-Ours}, the Jordan-Wigner transformation 
\eqref{JW-trans} 
is applied to $ \check{\tR}_{12}$. 
The similarity transformation by the matrix 
\eqref{shuffle} shuffles the matrix. 
This matrix ${\mathcal C}$ appeared on the right hand side of eq.\ (9) in \cite{Martins:2007hb}. 
In addition, the diagonal twist  (cf. \cite{Reshetikhin:1990ep})  by the matrix \eqref{signtwist}
changes the signs of some of its matrix elements. 
We remark that \eqref{eq:AFZ-Ours} satisfies the non-graded Yang-Baxter equation 
$S_{12}(p_{1},p_{2})S_{13}(p_{1},p_{3})S_{23}(p_{2},p_{3})=
S_{23}(p_{2},p_{3})S_{13}(p_{1},p_{3})S_{12}(p_{1},p_{2}) $ 
under the conditions \eqref{eq:massshellcondition} and \eqref{WS-momentum}. 

\subsection{Martins-Melo \cite{Martins:2007hb}}
The connection between the R-matrix in eq.\ (10) in  \cite{Martins:2007hb} 
and the above S-matrix in 
\cite{Arutyunov:2006yd} is summarized in eq. (8) in \cite{Martins:2007hb}. 
Under the relations \eqref{WS-momentum} and \eqref{AFZ-unitary}, 
it is given\footnote{We had to replace the function $a_{7}(\lambda, \mu) $ in 
eq.\ (10) in 
\cite{Martins:2007hb} with $-a_{7}(\lambda, \mu) $ 
to establish this connection \eqref{eq:MM-AFZ}. 
Note that the mass shell condition \eqref{eq:massshellcondition} 
is not required for \eqref{eq:MM-AFZ}.}
 by 
\begin{align}
R^{\text{MM}}_{12}&={\mathcal P}_{12} G_{1}G_{2} S_{12}(p_{1},p_{2}) G_{1}^{-1}G_{2}^{-1} , 
 \label{eq:MM-AFZ}
 \\
& \qquad G_{k}=(E_{11}+g_{k} (E_{22}+E_{33}) +E_{44} ){\mathcal C} , 
\nonumber 
\end{align}
where $ {\mathcal P}_{12} =\sum_{i,j=1}^{4} E_{ij} \otimes E_{ji}$, and 
the notations $\overline{R}_{12}(\lambda,\mu), 
x^{\pm} (\lambda), x^{\pm} (\mu), t(\lambda)$ and $t(\mu)$ in
 \cite{Martins:2007hb}
are expressed as $R^{\text{MM}}_{12}, x^{\pm}_{1} ,x^{\pm}_{2} ,g_{1}$ and $g_{2}  $, respectively. 
Substituting \eqref{eq:AFZ-Ours} into \eqref{eq:MM-AFZ}, 
we establish the connection between our two-layer S-matrix \eqref{eq:finalexpressionSmatrix} and the R-matrix in 
\cite{Martins:2007hb}. 
\subsection{Beisert \cite{Beisert:2006qh}}
\label{connection-Bei}
 First, we transcribe the S-matrix $ {\mathcal S}_{12}$ of Table 1 in \cite{Beisert:2006qh} 
 by interpreting his notations as follows. 
 We consider the case where the condition 
\begin{align}
\frac{ x^{+}_{k} }{ x^{-}_{k} } =\xi_{k}^{2}, 
\qquad k \in {\mathbb Z}_{\ge 1}, 
 \label{eq:symmetryconditions-app}
\end{align}
holds. 
This condition \eqref{eq:symmetryconditions-app} corresponds to the first equation in eq.\ (4.11) 
in \cite{Beisert:2006qh}.  
Under this condition, the contribution of the markers 
${\mathcal Z}^{\pm}, {\mathcal Y}^{\pm} $ can be neglected  
(see eq.\ (3.57) in \cite{Beisert:2006qh}). 
 Then we formally ignore the markers ${\mathcal Z}^{\pm}, {\mathcal Y}^{\pm} $, and 
  regard the vectors as
 $\ket{\Omega^{a}_{1} \Lambda^{b}_{2} } = \ket{\Omega^{a} } \otimes_{s} \ket{\Lambda^{b} } $, 
  $\ket{\Omega^{a}_{2} \Lambda^{b}_{1} } = \ket{\Omega^{a} } \otimes_{s} \ket{\Lambda^{b} } $, 
 where $\Omega ,\Lambda$ are $\phi$ or $\psi $, and $\otimes_{s}$ is 
 a graded (super) tensor product\footnote{See Appendix \ref{cone-gr-YBE} for $N=4$.}. 
 We set  $\ket{\phi^1}=e_1$, $\ket{\phi^2}=e_4$, $\ket{\psi^1}=e_2$, $\ket{\psi^2}=e_3$ 
 and assign the grading 
 $p(1)=p(e_{1})=p(4)=p(e_{4})=0$ and $p(2)=p(e_{2})=p(3)=p(e_{3})=1$, 
 where $e_i$ are canonical basis vectors, 
 which satisfy $E_{ij}e_{k}=\delta_{jk}e_{i}$. 
 Then the matrix elements $S^{i_{1}i_{2}}_{j_{1}j_{2}}$ 
 of the S-matrix $ {\mathcal S}_{12} $ are defined by\footnote{Do not confuse the matrix unit $E_{i_{1} j_{1}}$ with 
 the notation for the coefficient $E_{12}$.}
 \begin{align}
 {\mathcal S}_{12} (e_{j_{1}} \otimes_{s} e_{j_{2}}) 
& =
 \sum_{i_{1},i_{2}=1}^{4}  
  (e_{i_{1}} \otimes_{s} e_{i_{2}}) 
  S^{i_{1}i_{2}}_{j_{1}j_{2}}, 
 \\[6pt]
 \begin{split}
 & S^{11}_{11}=S^{44}_{44}=A_{12}, \quad S^{14}_{14}=S^{41}_{41}=\frac{A_{12}+B_{12} }{2}, 
 \quad S^{41}_{14}=S^{14}_{41}=\frac{A_{12}-B_{12} }{2}, 
 \\[5pt]
  &S^{23}_{14}=-S^{32}_{14}=S^{32}_{41}=-S^{23}_{41}=\frac{C_{12}}{2}, 
  \qquad S^{22}_{22}=S^{33}_{33}=D_{12}, 
\\[5pt]
& S^{23}_{23}=S^{32}_{32}=\frac{D_{12}+E_{12} }{2}, 
\qquad S^{32}_{23}=S^{23}_{32}=\frac{D_{12}-E_{12} }{2}, 
\\[5pt]
  &S^{14}_{23}=-S^{41}_{23}=-S^{14}_{32}=S^{41}_{32}=\frac{F_{12}}{2}, 
  \qquad  S^{21}_{12}=S^{31}_{13}=S^{24}_{42}=S^{34}_{43}=G_{12}, 
\\[5pt]
& S^{12}_{12}=S^{13}_{13}=S^{42}_{42}=S^{43}_{43}=H_{12}, 
\qquad  S^{21}_{21}=S^{24}_{24}=S^{31}_{31}=S^{34}_{34}=K_{12}, 
\\[5pt]
 & S^{12}_{21}=S^{42}_{24}=S^{13}_{31}=S^{43}_{34}=L_{12} . 
\end{split}
\nonumber 
\end{align}
Let us define an S-matrix ${\mathcal S}^{B}_{12}$ by\footnote{In contrast, ${\mathcal S}_{12}$ is realized by 
 $ {\mathcal S}_{12}=
 \sum_{i_{1},i_{2},j_{1},j_{2}=1}^{4}  
 (-1)^{p(j_{1})(p(i_{2}) +p(j_{2}) ) }
  S^{i_{1}i_{2}}_{j_{1}j_{2}} \ 
  E_{i_{1}j_{1}} \otimes_{s}  E_{i_{2}j_{2}} $ 
 on the ${\mathbb Z}_{2}$-graded vector space. 
 Let ${\mathcal P}^{g}_{12}$ be the graded permutation 
 $ {\mathcal P}^{g}_{12}=\sum_{i,j} (-1)^{p(j)}E_{ij} \otimes_{s} E_{ji}$. 
  The R-matrix ${\mathcal R}^{g}_{12}={\mathcal P}^{g}_{12}  {\mathcal S}_{12} $ 
 satisfies the graded Yang-Baxter equation 
 $
 {\mathcal R}^{g}_{12}{\mathcal R}^{g}_{13}{\mathcal R}^{g}_{23}=
 {\mathcal R}^{g}_{23}{\mathcal R}^{g}_{13}{\mathcal R}^{g}_{12}
 $
 under the relation \eqref{eq:symmetryconditions-app} and 
 the mass shell condition \eqref{eq:massshellcondition}. 
 See Appendix \ref{cone-gr-YBE}.
}
\begin{align}
  {\mathcal S}^{B}_{12}&=
 \sum_{i_{1},i_{2},j_{1},j_{2}=1}^{4}  
  S^{i_{1}i_{2}}_{j_{1}j_{2}} \ 
  E_{i_{1}j_{1}} \otimes  E_{i_{2}j_{2}} .
\end{align}
We remark that 
 the R-matrix ${\mathcal R}^{B}_{12}={\mathcal P}_{12}  {\mathcal S}^{B}_{12} $ 
 satisfies the non-graded Yang-Baxter equation 
 $
 {\mathcal R}^{B}_{12}{\mathcal R}^{B}_{13}{\mathcal R}^{B}_{23}=
 {\mathcal R}^{B}_{23}{\mathcal R}^{B}_{13}{\mathcal R}^{B}_{12}
 $
 under 
 the relation \eqref{eq:symmetryconditions-app} and 
 the mass shell condition \eqref{eq:massshellcondition}. 
Then, using the variable identification 
\begin{align}
x^{+}_{k}= \frac{\Theta}{\Xi} v_{k},  
 \quad 
x^{-}_{k}= \frac{\Theta}{\Xi}
              \frac{b_{k} c_{k} }{a_{k} d_{k}}  v_{k},
         \quad  
 \xi_{k}= - \frac{d_{k}}{b_{k}},
  \quad 
 \frac{\gamma_{k}}{ \sqrt{ -i \alpha}}
=\frac{\sqrt{x^{+}_{k}}}{a_{k}} 
 \left (-\frac{a_{k} d_{k}}{b_{k} c_{k}} \right)^{\frac{1}{4}}   , 
 \quad  
 g=-\Theta \Xi , 
  \label{eq:variablesidentifications-app}
\end{align}
which is 
similar to \eqref{eq:variablesidentifications}, one can relate the  Shastry-Shiroishi-Wadati ${\mathfrak R}_{12}$
 \eqref{eq:generalRmatrixSpin}
 to ${\mathcal S}^{B}_{12}$. 
 Here the parameters $\xi_{k}, \gamma_{k}, \alpha $ are the ones in \cite{Beisert:2006qh}. 
With this identifications \eqref{eq:variablesidentifications-app}, the gluing condition \eqref{eq:gluingcond} becomes the mass shell condition:
\eqref{eq:massshellcondition}.
One can invert the first 4 relations in \eqref{eq:variablesidentifications-app} as
\begin{align}
&a_{k}=\frac{\sqrt{-i \alpha} }{\gamma_{k}} \sqrt{x^{+}_{k} } \left( -\frac{x^{+}_{k} }{x^{-}_{k}} \right)^{\frac{1}{4}} ,& 
&b_{k}=-\frac{1}{\xi_{k} a_{k}} \frac{x^{+}_{k}}{ x^{+}_{k} -x^{-}_{k} }, &
&c_{k}=-\xi_{k} a_{k} \frac{ x^{-}_{k} }{ x^{+}_{k} }, &
\nonumber\\
&d_{k}=\frac{1}{ a_{k} } \frac{ x^{+}_{k} }{ x^{+}_{k} -x^{-}_{k} } ,&
&v_{k}= \frac{\Xi}{\Theta} x^{+}_{k} .&
\label{eq:variablesidentifications-inv-app}
\end{align}
Note that the symmetry conditions 
\eqref{eq:symmetryconditions} are equivalent to \eqref{eq:symmetryconditions-app} 
under \eqref{eq:variablesidentifications-inv-app}. 
 We introduce the matrices 
\begin{align}
W & \colonequals \sigma^{z} \otimes 
\text{diag}(1,-i,-i,1)\otimes\mathbbm{1}_2, 
\nonumber\\[6pt]
U_{k}&\colonequals  
\mathrm{diag}
\left(
\left(- \frac{ x^{-}_{k} }{x^{+}_{k} }  \right)^{\frac{1}{4}} \sqrt{ x^{+}_{k} }, 1,1,
-
\left(- \frac{ x^{+}_{k} }{x^{-}_{k} }  \right)^{\frac{1}{4}} \sqrt{ x^{+}_{k} } 
\frac{ i\alpha }{\gamma_{k}^{2} }  ( x^{+}_{k} -x^{-}_{k} )
\right),  
\end{align}
for $k=1,2.$
Then we  find that the following relation  
\beq
\label{eq:relSRbos}
\frac{{\mathcal S}^{B}_{12}}{S^{0}_{12}}=
\rho_{12} 
W {\mathcal P}_{12}
\big(U_1\otimes U_2 \big)\bar{P}_{12}
{\mathfrak R}_{12}
\bar{P}_{12}\big(U_1^{-1}\otimes U_2^{-1}\big)W^{-1}, 
\eeq
holds under the symmetry conditions \eqref{eq:symmetryconditions-app}, 
where $\rho_{12}$ is an overall factor. 
An explicit expressions for $\rho_{12}$ is given by 
\begin{align}
\rho_{12} 
=
\frac{b_{1}d_{2}(a_{2}d_{2}v_{1}+b_{2}c_{2}v_{2})}
{ b_{2}d_{1}(a_{1}a_{2}d_{1}d_{2}-b_{1}b_{2}c_{1}c_{2})
(b_{2}c_{2}v_{2}-a_{2}d_{2}v_{1})}
= \frac{ \xi_{2} ( x^{+}_{1} - x^{-}_{1} ) ( x^{+}_{2} - x^{-}_{2} ) ( x^{-}_{2} + x^{+}_{1} ) }
{ \xi_{1} ( x^{+}_{1} x^{+}_{2} - x^{-}_{1} x^{-}_{2} ) ( x^{-}_{2} - x^{+}_{1} ) }
  . 
\end{align}
The matrix $W$ was introduced following \cite{Akutsu:1987} to account for the difference of grading, since ${\mathcal S}_{12}$ is graded, while the Shastry-Shiroishi-Wadati R-matrix of \eqref{eq:generalRmatrixSpin} isn't. 
\section{The double free fermion condition}
\label{double-free}
\addcontentsline{toc}{section}{Appendix D}
\def\theequation{D\arabic{equation}}
\setcounter{equation}{0}
In this appendix, we would like to present another aspect of the connection between the Shastry-Shiroishi-Wadati R-matrix of \eqref{eq:generalRmatrixfermionic} and the AdS/CFT S-matrix. As pointed out in \cite{Bazhanov:1984iw}, the R-matrix of the free fermion model satisfies the condition \eqref{eq:parametrizeSL2C}.  We want to introduce a generalization of this condition, the double free fermion condition,  and to write it for the AdS/CFT S-matrix. 
On the linear space of $4\times 4$ complex matrices, we can introduce
\footnote{Here $R_{ij}$ denotes the $(i,j)$-matrix element of a 
$4 \times 4$ matrix $R$. It should not be confused
 with $R_{ij}$, which is used in the other part of the text, 
to denote $R$ acting  
on the lattice site labeled by $(i,j)$.}
 the following bilinear form:
\beq
\left(R,R^{\prime}\right)\colonequals R_{11}R^{\prime}_{44}+R_{44}R^{\prime}_{11}+R_{22}R^{\prime}_{33}+R_{33}R^{\prime}_{22}-R_{23}R^{\prime}_{32}-R_{32}R^{\prime}_{23}\ .
\eeq
Then we find that the matrices $R^i$ satisfy the relations
\beq
\left(R^0,R^0\right)=\left(R^1,R^1\right)=\left(R^0,R^1\right)=0\ .
 \label{eq:bilinear}
\eeq
The first two follow from $\det A_1=\det A_2=1$. We refer to the third relation as the \textit{compatibility equation}. It 
does not depend on the condition that the determinant of the matrices $A_1$ and $A_2$ is one, but follows instead from the specific form of the $R^i$ matrices. Let us now introduce a $16\times 16$-dimensional matrix   $\tR = \sum_{i,j,k,l=1}^4 \tR^{ik}_{\phantom{ik}jl}  E_{ij} \otimes E_{kl}$, where 
$E_{ab}$ is a $4\times 4$ matrix unit.  
We require that $\tR$ has only $36$ non-vanishing entries by imposing:
\beq
\label{eq:vanish36}
\tR^{ik}_{\phantom{ik}jl}=0\text{ if } (i,j)\text{ or }(k,l) \notin \{(1,1),(2,2),(2,3),(3,2),(3,3),(4,4)\}\ .
\eeq
We also require that $\tR$ satisfies the free fermion 
condition both in the first space and the second space in the tensor product.  
In components, this can be written as
\beq
\label{eq:doublefreefermion}
\tR^{i1}_{\phantom{i1}j1}\tR^{i4}_{\phantom{i4}j4}+
\tR^{i2}_{\phantom{i2}j2}\tR^{i3}_{\phantom{i3}j3}-
\tR^{i2}_{\phantom{i2}j3}\tR^{i3}_{\phantom{i3}j2}=
\tR^{1k}_{\phantom{1k}1l}\tR^{4k}_{\phantom{4k}4l}+
\tR^{2k}_{\phantom{2k}2l}\tR^{3k}_{\phantom{3k}3l}-
\tR^{2k}_{\phantom{2k}3l}\tR^{3k}_{\phantom{3k}2l}=0\ .
\eeq
We call this condition the {\em double free fermion condition}.  
Taken together, \eqref{eq:bilinear} imply that any matrix of the form
\beqa
\label{eq:genansatzfreeferm}
\tR&=&\sum_{r,s=0}^1c_{rs} R^r(A_1,A_2)\otimes R^s(A_3,A_4),
\eeqa
with $A_r\in \text{SL}(2,\mathbb{C})$, satisfies the double free fermion condition for any value of the complex parameters $c_{rs}$. Since the Shastry-Shiroishi-Wadati R-matrix of \eqref{eq:generalRmatrixfermionic}, written here using matrices instead of oscillators, is precisely of the form \eqref{eq:genansatzfreeferm}, we know that the double free fermion condition is obeyed. We would now like to find the implications of this for the AdS/CFT correspondence.
Inserting ${\mathfrak R}$ into \eqref{eq:doublefreefermion} and making use of \eqref{eq:relSRbos} to express everything in the coefficients of ${\mathcal S}_{12}$, 
leads to the following\footnote{See Table 1 in \cite{Beisert:2006qh} for the notation.} three quadratic equations:
\beqa
\label{eq:doubleffBeisert}
A_{12}D_{12}&=&H_{12}K_{12}-G_{12}L_{12} , \nonumber\\
B_{12}E_{12}-C_{12}F_{12}&=&H_{12}K_{12}-G_{12}L_{12} , \nonumber\\
A_{12}E_{12}+B_{12}D_{12}&=&2\left(H_{12}K_{12}+G_{12}L_{12}\right)\ .
\eeqa
It turns out that the first two equations are to be found in the article \cite{Beisert:2008tw}, where they were pointed out in eq.\ (3.6) as a curious occurrence in the context of unitarity of the $R$ matrix. The third one on the other hand follows for the others by making use of the mass-shell condition \eqref{eq:massshellcondition}. Here, we derive these three equations as a direct consequence of the two-layer structure of the model. 


\end{document}